\documentclass[%
    reprint,
    superscriptaddress,
    amsmath,amssymb,
    aps,
]{revtex4-2}

\usepackage{graphicx}
\usepackage{dcolumn}
\usepackage{bm}
\usepackage{algorithm}
\usepackage{algorithmic}
\usepackage{caption}
\usepackage{ragged2e}

\usepackage{booktabs}
\usepackage{subcaption}
\usepackage{tikz}
\usepackage{enumitem}
\usepackage{siunitx}
\usepackage{placeins}
\usepackage{dblfloatfix}

\begin{document}

\preprint{APS/123-QED}

\title{Refined Gradient-Based Temperature Optimization for\\ the Replica-Exchange Monte-Carlo Method}

\author{Tatsuya Miyata}
\affiliation{Institute of Science Tokyo, Ookayama, Tokyo 152-8550, Japan}
\author{Shunta Arai}
\affiliation{Institute of Science Tokyo, Ookayama, Tokyo 152-8550, Japan}
\author{Satoshi Takabe}
\affiliation{Institute of Science Tokyo, Ookayama, Tokyo 152-8550, Japan}
\affiliation{RIKEN Center for Advanced Intelligence Project (AIP), Nihonbashi, Tokyo 103-0027, Japan}

\date{\today}

\begin{abstract}
The replica-exchange Monte-Carlo (RXMC) method is a powerful Markov-chain Monte-Carlo algorithm for sampling from multi-modal distributions, which are challenging for conventional methods. 
The sampling efficiency of the RXMC method depends highly on the selection of the temperatures, and finding optimal temperatures remains a challenge. 
In this study, we propose a refined online temperature selection method by extending the gradient-based optimization framework proposed previously.
Building upon the existing temperature update approach, we introduce a reparameterization technique to strictly enforce physical constraints, such as the monotonic ordering of inverse temperatures, which were not explicitly addressed in the original formulation.
The proposed method defines the variance of acceptance rates between adjacent replicas as a loss function, estimates its gradient using differential information from the sampling process, and optimizes the temperatures via gradient descent.
We demonstrate the effectiveness of our method through experiments on benchmark spin systems, including the two-dimensional ferromagnetic Ising model, the two-dimensional ferromagnetic XY model, and the three-dimensional Edwards-Anderson model. Our results show that the method successfully achieves uniform acceptance rates and reduces round-trip times across the temperature space.
Furthermore, our proposed method offers a significant advantage over a recently proposed policy gradient method that requires careful hyperparameter tuning, while simultaneously preventing the constraint violations that destabilize optimization.
\end{abstract}

\maketitle

%\tableofcontents

\section{\label{sec:level1}Introduction}

Markov-chain Monte-Carlo (MCMC) methods~\cite{Metropolis1953, Hastings1970} are standard tools for sampling from complex probability distributions. 
However, conventional MCMC methods often encounter difficulties when sampling from multi-modal distributions, particularly the Boltzmann distributions of systems with rugged energy landscapes. 
In such systems, the Markov chain can become trapped in local energy minima, leading to insufficient exploration of the state space and slow relaxation.

The replica-exchange Monte-Carlo~(RXMC) method~\cite{Hukushima1996, Swendsen1986, Geyer1991}, also known as parallel tempering, is widely used to alleviate this problem.
In the RXMC method, multiple Markov chains, known as replicas, are simulated at different temperatures, and their states are periodically exchanged between adjacent replicas according to an acceptance rate that satisfies the detailed balance condition.
High-temperature replicas can easily overcome free energy barriers by exploring a flattened energy landscape, while low-temperature replicas perform precise local exploration. 
The exchange mechanism induces a random walk in temperature space, allowing replicas to escape local minima and reach different modes.
The RXMC method is widely applied to spin glass~\cite{Hukushima1996,Swendsen1986} and protein folding~\cite{Hansmann1997}.

The sampling efficiency of the RXMC method depends critically on the selection of the temperatures.
Inappropriate temperature selection can create a bottleneck where the acceptance rate between specific adjacent replicas becomes very small, thereby hindering the random walk in temperature space.
A common heuristic strategy is the geometric progression~\cite{Predescu2004}. This achieves uniform acceptance rates under the assumption of constant specific heat. 
However, in systems exhibiting phase transitions, this assumption breaks down because the specific heat has a sharp peak near the critical point, leading to a serious decrease in sampling efficiency.
% Therefore, establishing a scheme to automatically determine the optimal temperature without prior knowledge for target systems with complex energy landscapes and phase transitions is valuable.

% 4. Existing Approaches and Limits
To address these problems, various temperature selection methods have been proposed. 
However, conventional heuristic methods still face practical difficulties.
For example, offline optimization methods that attempt to determine an optimal temperature configuration in advance require preliminary sampling to estimate quantities such as the system's specific heat and the density of states~\cite{Hukushima1996, Hukushima1999, Sugita1999, Kofke2002, Rathore2004, Kone2005, Katzgraber2006, Gront2007, Nadler2007a, Denschlag2009, Miasojedow2013, Vousden2015, Rozada2019, Syed2021}.
This problem becomes particularly severe near a critical point, where the specific heat exhibits a sharp peak.
In such a region, even a small error in the temperature setting can substantially reduce the acceptance rates.
Another class of approaches consists of metaheuristic methods such as evolutionary strategies~\cite{Kowaguchi2022}.
By exploring multiple candidate temperature vectors, these methods can help avoid stagnation in local optima of the temperature optimization objective function.
However, they typically require the evaluation of many candidates, often in parallel, and therefore demand more computational resources, such as computation time and memory, as the number of candidates increases.
In addition, their performance depends on hyperparameters such as the perturbation variance and learning rate, which must be tuned carefully under noisy estimates of sampling efficiency.
Therefore, it is useful to develop a more versatile online approach that optimizes the temperature during simulation while reducing preliminary calculations and manual tuning.

% 5. Gradient-based Approaches
Recently, gradient-based online optimization methods have emerged as a promising alternative~\cite{MacCallum2018, Zhao2024}. 
These approaches define a differentiable loss function related to sampling efficiency and update the inverse temperature vector using gradient estimates obtained from statistics collected during simulation.
The advantage of this approach is that it can optimize within a simulation run without detailed prior knowledge of the model.
However, existing gradient-based methods still have several challenges. 
Methods that directly calculate the gradient of the loss function do not explicitly handle the monotonic constraint inherent in the inverse temperature vector, which leads to instability in the optimization process. 
In reinforcement learning approaches~\cite{Zhao2024}, the parameterization of the policy results in the burden of new hyperparameter tuning. 
Furthermore, the effectiveness of these gradient-based methods has mainly been examined in continuous-variable systems, and their behavior in discrete spin systems with phase transitions has not been fully investigated.

% 6. Our Contribution
Therefore, this study aims to develop a refined online temperature selection method by extending the gradient-based optimization framework proposed previously~\cite{MacCallum2018}.
While adopting their temperature update approach, we introduce a reparameterization technique to address monotonicity constraints that were not explicitly enforced in their original formulation.
Our contributions are as follows:
\begin{enumerate}[label=\arabic*), topsep=1pt, itemsep=0pt]
    \item We refine the optimization framework in~\cite{MacCallum2018} by introducing a reparameterization based on inverse temperature differences.
    This formulation incorporates the monotonic ordering of inverse temperatures and fixed boundary conditions into the optimization variables, ensuring that these constraints are satisfied during optimization.
    It also avoids the additional policy hyperparameter tuning required in reinforcement learning approaches.
    \item We demonstrate the effectiveness of the proposed method on spin systems exhibiting phase transitions, where the applicability of conventional gradient-based methods remains unclear. 
    Specifically, we apply it to the Ising model, the XY model, and the Edwards-Anderson model, showing that it can achieve uniform acceptance rates between adjacent replicas.
\end{enumerate}

The remainder of this paper is organized as follows. 
Section~\ref{sec:level2} reviews related work, and Section~\ref{sec:level3} details our proposed method. 
Section~\ref{sec:level4} presents the experimental setup, results, and discussion. 
Finally, Section~\ref{sec:level5} concludes the paper.

\section{\label{sec:level2}Related Work}
This section reviews the background and prior work relevant to the proposed method. 
We first provide a brief overview of the RXMC method and then introduce the round-trip time as a metric for evaluating its efficiency. 
We discuss recent temperature optimization methods based on policy gradient methods, highlighting their strengths and limitations.

\subsection{Replica-exchange Monte-Carlo}

In the RXMC method, $M$ non-interacting copies of the target system, called replicas, are simulated simultaneously with a monotonically ordered inverse temperature vector $\bm\beta = (\beta_1, \beta_2, \dots, \beta_M)$, where $\beta_1 < \beta_2 < \dots < \beta_M$.
The lowest inverse temperature $\beta_1$ is set small enough to allow the system to overcome large energy barriers, while the highest inverse temperature $\beta_M$ is set large enough to sample local energy minima in detail.
The RXMC method consists of two alternating processes.
The first is the intra-replica update.
Each replica is updated independently for a specified number of Monte-Carlo steps (MCS).
The update is performed using a standard MCMC procedure, such as the Metropolis--Hastings algorithm~\cite{Hastings1970}, according to the target distribution at that replica's assigned inverse temperature.
This allows each replica to explore the state space locally and sample its own canonical ensemble.
The second is the inter-replica exchange, which characterizes the RXMC method. Following the intra-replica update, exchanges are attempted between replicas at adjacent inverse temperatures.
For a pair of replicas at neighboring inverse temperatures $\beta_i$ and $\beta_{i+1}$ ($i=1,\dots,M-1$), a proposal is made to swap their states.
This proposed swap is accepted with probability based on the detailed balance condition for the extended ensemble.
The acceptance rate between adjacent replicas is given by the Metropolis criterion:
\begin{equation}
    A_{i,i+1}(\beta_i, \beta_{i+1}) = \min \left\{ 1, \exp(\Delta\beta \Delta E ) \right\},
    \label{eq:remc_swap}
\end{equation}
where $\Delta\beta = \beta_{i+1} - \beta_i$ is the difference between the inverse temperatures and $\Delta E = E_{i+1} - E_i$ is the energy difference between the states of the two replicas.

The advantage of RXMC comes from this exchange mechanism.
Conventional MCMC methods based on local updates often become trapped in local minima separated by energy barriers.
In contrast, the swap moves in RXMC induce a random walk in temperature space, enabling replicas to move to high temperature regions where equilibration is faster and subsequently return to low temperature regions.
This mechanism helps the system cross energy barriers and reduces relaxation times.

The efficiency of this exchange process relies heavily on the choice of the inverse temperature vector.
For the random walk in temperature space to proceed smoothly, there must be sufficient overlap between the Boltzmann distributions of adjacent replicas to maintain non-vanishing acceptance rates between adjacent replicas.
A widely used heuristic for determining this placement is the geometric progression~\cite{Predescu2004}, which arranges inverse temperatures according to
\begin{equation}
    \beta_j = \beta_M \left( \frac{\beta_1}{\beta_M} \right)^{\frac{M-j}{M-1}}\quad (j=1,2,\dots, M).
    \label{eq:geometric}
\end{equation}
This strategy gives denser temperature intervals in the low-temperature region, where relaxation is assumed to be slower.
It can achieve uniform acceptance rates between adjacent replicas when the system's specific heat is constant regardless of temperature~\cite{Kofke2002}.
However, this assumption does not generally hold, especially in systems exhibiting phase transitions.
As a result, the acceptance rates become unbalanced, producing bottlenecks in certain temperature ranges.
To address this issue, various heuristic methods have been developed to determine system-dependent temperatures, as reviewed in the previous section.

\subsection{Round-trip Time}
% \begin{figure}[t]
% \includegraphics[width=0.5\textwidth]{fig/round_trip_diagram.png}
% \caption{This illustrates the movement of a replica through the inverse temperature space during the simulation. Each replica traverses the entire simulated range $[\beta_{\mathrm{min}}, \beta_{\mathrm{max}}]$. A full round-trip consists of one process ascending from $\beta_{\mathrm{min}}$ to $\beta_{\mathrm{max}}$ duration $\tau_{\mathrm{up}}$ and one process descending back to $\beta_{\mathrm{min}}$ duration $\tau_{\mathrm{down}}$. The average round-trip time, $\tau_{\mathrm{rt}}$, is the sum of these two durations: $\tau_{\mathrm{rt}} = \tau_{\mathrm{down}} + \tau_{\mathrm{up}}$.}
% \label{fig:round-trip_time_diagram}
% \end{figure}

A useful metric for evaluating the sampling efficiency of the RXMC method is the round-trip time $\tau_{\mathrm{RT}}$.
It is defined as MCS required for a single replica to travel back and forth between the lowest and highest inverse temperatures.

The round-trip time is conceptually divided into two parts.
Consider a replica starting at the highest inverse temperature $\beta_M$.
The first part, $\tau_{\mathrm{down}}$ is the time taken for this replica to perform a random walk in the inverse temperature space and reach the lowest inverse temperature $\beta_1$ for the first time.
The second part, $\tau_{\mathrm{up}}$ is the subsequent time taken for the replica to return to $\beta_M$ for the first time.
The round-trip time is thus defined as $\tau_{\mathrm{RT}} = \tau_{\mathrm{down}} + \tau_{\mathrm{up}}$. 
A shorter $\tau_{\mathrm{RT}}$ indicates more efficient movement of replicas through inverse temperature space.
It reflects frequent transitions in which configurations sampled at low temperatures are relaxed at high temperatures and then returned to the low-temperature region.
Thus, a shorter round-trip time implies more efficient exploration of the state space.

To measure these times in a simulation, a labeling scheme can be employed~\cite{Katzgraber2006}.
Each replica is assigned a label indicating its direction of travel in temperature space.
For instance, a replica at $\beta_M$ is assigned the label ``down''. 
This label remains until the replica reaches the opposite end $\beta_1$, for the first time, at which point its label changes to ``up''.
Similarly, a replica with the label ``up'' changes its label back to ``down'' when it returns to $\beta_M$ for the first time.
By recording the time intervals between these label changes, $\tau_{\mathrm{down}}$ and $\tau_{\mathrm{up}}$ are measured, and the round-trip time $\tau_{\mathrm{RT}}$ is then obtained.

\subsection{\label{subsec:2.C}Policy Gradient Method for Temperature Selection}

The temperature selection problem in the RXMC method has been modeled as a single-state reinforcement learning problem, applying the policy gradient method~\cite{Zhao2024}.
In this context, a single-state problem implies that the environment does not transition between different states based on the agent's actions.

In their approach, an action $\bm a$ representing the temperatures was sampled from a parameterized stochastic policy $\pi_{\bm\theta}$, and the policy parameters $\bm\theta$ were updated based on a reward quantifying the sampling efficiency.

The objective in policy gradient methods is to find the optimal parameters $\bm\theta$ that maximize the expected reward $J(\bm\theta) := \mathbb{E}_{\pi_{\bm\theta}}[R_{\bm a}]$, where $R_{\bm a}$ is the reward obtained for the action $\bm a$.
The parameters are optimized via gradient ascent: 
% The policy gradient theorem provides an expression for the gradient of the expected reward with respect to the parameters. For the single-state case, this gradient is typically estimated with a Monte-Carlo approach: an action $\bm a_t$ is sampled from the current policy $\pi_{\bm\theta_t}$, the sampler is run to observe a reward $r_t$, and the parameters are updated according to the following rule:
\begin{equation}
    \bm\theta \leftarrow \bm\theta + \eta (J(\bm\theta) - \bar{R}) \nabla_{\bm\theta} \log \pi_{\bm\theta}(\bm a),
\end{equation}
where $\eta$ is the learning rate, and $\bar{R}$ is a reward baseline used to reduce the variance of the gradient estimate.
Algorithm~\ref{alg1} outlines the policy gradient method in the single-state case.

\begin{algorithm}[H]
	\caption{Single-State Policy Gradient~\cite{Zhao2024}}
	\label{alg1}
	\begin{algorithmic}[1]
        
    \REQUIRE policy distribution $\pi_{\bm\theta}(\cdot)$, learning rate $\eta$, epochs $T$
    \item[\textbf{Initialize:}] policy parameters $\bm\theta_0$, MCMC states $\{\bm x_0^{(i)}\}_{i=1}^M$

    \FOR{$t = 0, 1, \dots, T-1$}
        \STATE Generate a vector of sampler parameters $\bm a_t$ by sampling from the policy distribution $\pi_{\bm\theta_t}$.
        \STATE Run the RXMC sampler with parameters $\bm a_t$ and observe the reward $R_{\bm a_t}$.
        \STATE Update the average reward $\bar{R}$ based on observed rewards.
        \STATE Compute the gradient of policy $g_t\!=\!\nabla_{\bm\theta} \log \pi_{\bm\theta}(\bm a_t)|_{\bm\theta=\bm\theta_t}$ w.r.t. $\bm\theta$ and clip it.
        \STATE Update the policy parameters: $\bm\theta_{t+1} \!=\! \bm\theta_t + \eta (R_{\bm a_t} - \bar{R}) g_t$.
    \ENDFOR

    \ENSURE Final optimized policy parameters $\bm\theta_T$.
	\end{algorithmic}
\end{algorithm}
Specifically, the action space was defined using the differences of logarithmic inverse temperatures $D_i = \log\beta_{i+1} - \log\beta_i$, resulting in $\bm a = (D_1, \dots, D_{M-1}) \in [0, \infty)^{M-1}$. 
Using the differences of logarithmic inverse temperatures ensures that the scale of all parameters is similar, which contributes to learning stability and allows for easily fixing the boundary conditions. 
A multivariate Gaussian distribution with mean $\bm\theta$ and variance-covariance matrix $\sigma^2 \bm{I}$ was used as the policy distribution, where $\bm{I}$ denotes the identity matrix. 
The primary reason for this choice is the analytical simplicity of the gradient computation.
Regardless of the complexity of the target system, this formulation reduces the gradient of the log-policy to a simple form:
\begin{equation}
    \nabla_{\bm\theta} \log \pi_{\bm\theta}(\bm a_t) = \frac{\bm a_t - \bm\theta}{\sigma^2}.
\end{equation}

However, the experiments in \cite{Zhao2024} were conducted only for continuous distributions, such as the multi-modal Gaussian distribution and Boltzmann distributions with Rosenbrock and egg-box functions. 
Thus, the applicability of their method to discrete variable systems, particularly spin systems such as the Ising model, remains unveiled. 
Specifically, the assumption that the differences of logarithmic inverse temperatures have a similar scale may not hold in systems exhibiting phase transitions because the required temperature intervals can vary significantly. 
This issue may destabilize the learning process.
Therefore, the effectiveness of the method needs to be verified for such systems.
Moreover, the method requires setting several hyperparameters such as the initial policy parameters $\bm\theta_0$, the policy variance $\sigma^2$, and the learning rate $\eta$. The tuning process can be complex and tedious. 
In particular, the policy variance $\sigma^2$ is critical to the convergence of the policy gradient method, and the method may fail to converge if it is not tuned appropriately.

\section{\label{sec:level3}Temperature Optimization}

In this study, we propose a refined framework to optimize the inverse temperatures in the RXMC method.
The core idea is to iteratively update the inverse temperature vector $\bm{\beta}$ to minimize a predefined loss function $f(\bm{\beta})$ that quantifies sampling efficiency. This optimization is performed using a gradient-based approach. 
A key feature of our framework is that the gradient of the loss function is estimated directly by utilizing observables obtained from the sampling process itself.

\subsection{\label{subsec:3.A}Loss Function}

Our primary objective is to achieve uniform acceptance rates between adjacent replicas.
This goal is motivated by the intuition that any local suppression of acceptance rates acts as a bottleneck, impeding the random walk of replicas in the temperature space.
Reducing such bottlenecks is expected to improve the movement of replicas across temperatures and shorten the relaxation time.
This goal of uniformity can be formulated as a minimization problem. 
Accordingly, the loss function is given by the variance of the expected acceptance rates across all adjacent replica pairs:
\begin{equation}
    f_{\mathrm{uni}}(\bm{\beta}) = \frac{1}{M-1} \sum_{i=1}^{M-1} \left( \mathbb{E}[A_{i,i+1}(\beta_i, \beta_{i+1})] - \bar{A} \right)^2,
\end{equation}
where $\mathbb{E}[\cdot]$ denotes the expectation taken with respect to the joint distribution of the relevant replicas, $A_{i,i+1}$ is the acceptance rate between adjacent replicas at inverse temperatures $\beta_i$ and $\beta_{i+1}$ as defined in Eq.~\eqref{eq:remc_swap} and $\bar{A} = \sum_{i=1}^{M-1} \mathbb{E}[A_{i,i+1}]/(M-1)$~\cite{MacCallum2018,Kowaguchi2022}. 
This loss function measures the non-uniformity of the expected acceptance rates between adjacent replicas.
Minimizing $f_{\mathrm{uni}}$ makes each expected acceptance rate close to the average value $\bar{A}$.
Thus, $f_{\mathrm{uni}}$ directly corresponds to our main objective achieving uniform acceptance rates between adjacent replicas.

As a related objective, a metric that maximizes the product of expected acceptance rates was proposed by MacCallum \textit{et al.}~\cite{MacCallum2018}. 
To apply gradient descent, this objective is reformulated as the sum of the negative natural logarithms of the expected acceptance rates:
\begin{equation}
    f_{\mathrm{log}}(\bm{\beta}) = - \sum_{i=1}^{M-1} \log \left( \mathbb{E}[A_{i,i+1}(\beta_i, \beta_{i+1})] \right).
\end{equation}
Since the negative logarithm assigns a large penalty to small acceptance rates, minimizing $f_{\mathrm{log}}$ encourages particularly small acceptance rates to increase.
Thus, $f_{\mathrm{log}}$ focuses on raising low acceptance rates, whereas $f_{\mathrm{uni}}$ focuses on equalizing acceptance rates between adjacent replicas.
In this study, we also assess the effectiveness of this metric on the Ising model and compare it with $f_{\mathrm{uni}}$. 
We utilize the gradient derivation provided in their work for this optimization.

While other metrics, such as the round-trip time or the integrated autocorrelation time, could be used as the loss function, they are not practical for online optimization. 
Estimating these quantities accurately requires large MCS, making them computationally expensive to evaluate at each step of an iterative optimization process. 
On the other hand, estimates obtained from short runs suffer from large statistical variance, which would make the learning process unstable. 
Therefore, we use $f_{\mathrm{uni}}$ as the primary loss function for temperature optimization in this study.
The metric $f_{\mathrm{log}}$ is used only as a comparative metric to examine how the choice of loss function affects the optimization behavior in Sec.~\ref{sec:Ising_model}.

\subsection{Gradient Estimation via the Score Function Method}

To minimize the loss function $f_{\mathrm{uni}}(\bm{\beta})$ defined in the previous section, we employ gradient descent. In this study, the minimum and maximum inverse temperatures $\beta_1$ and $\beta_M$ are fixed to ensure the inverse temperatures span the desired regimes. Consequently, the optimization targets the intermediate inverse temperatures from $\beta_2$ to $\beta_{M-1}$.

A straightforward approach is to treat the inverse temperature vector $\bm{\beta}$ directly as the parameters and compute its gradient $\nabla_{\bm{\beta}}f_{\mathrm{uni}}$. 
However, it is suggested that direct optimization in this parameter space can be unstable and prone to trapping in undesirable local minima~\cite{MacCallum2018}. 
Indeed, in our preliminary experiments described in Appendix~\ref{app:direct_optimization}, we observed that directly updating $\bm{\beta}$ often resulted in the stagnation of the optimization process.
The instability is primarily attributed to the characteristics of the gradient.
In the initial phase of optimization, the inverse temperatures tend to drift away from critical regions, leading to excessive intervals between adjacent replicas.
Consequently, the acceptance rates associated with these excessive intervals drop to nearly zero. In this regime, the optimization process loses the effective driving force to correct the intervals, leading to stagnation.
While employing a larger learning rate could help the system escape such stagnation, it inevitably leads to a violation of the monotonicity constraint $\beta_{1} < \beta_{2} < \dots < \beta_{M}$, as the update magnitude can easily exceed the interval between neighbors.

To circumvent these issues, we introduce a reparameterization of the variables. We adopt the logarithm of the inverse temperature differences as the optimization parameters. Specifically, we define the parameter vector $\bm{L} = (L_1, \dots, L_{M-2})$, where each element corresponds to the log-difference $L_k = \log\Delta\beta_k=\log(\beta_{k+1} - \beta_k)$ for $k=1, \dots, M-2$. 
Although there are $M-1$ intervals in total, the constraint of fixed endpoints imposes the condition $\sum_{k=1}^{M-1} \Delta\beta_k = \beta_M - \beta_1$. This constraint reduces the degrees of freedom by one, leaving $M-2$ independent variables to be optimized. 
This formulation allows for flexible exploration of the parameter space while strictly maintaining the order of temperatures.

The partial derivative of the loss function $f_{\mathrm{uni}}$ with respect to the parameter $L_k$ is derived using the chain rule as follows:
\begin{equation}
\begin{split}
    \frac{\partial f_{\mathrm{uni}}}{\partial L_k} &= \sum_{j=2}^{M-1}\frac{\partial f_{\mathrm{uni}}}{\partial \beta_j}\frac{\partial\beta_j}{\partial L_k} \\
    &= \sum_{j=2}^{M-1}\sum_{n=1}^{M-2}\frac{\partial f_{\mathrm{uni}}}{\partial \beta_j}\frac{\partial\beta_j}{\partial\Delta\beta_n}\frac{\partial\Delta\beta_n}{\partial L_k} \\
    &= \Delta\beta_k\sum_{j=2}^{M-1}\sum_{n=1}^{M-2}\frac{\partial f_{\mathrm{uni}}}{\partial \beta_j}\frac{\partial\beta_j}{\partial\Delta\beta_n}.
\end{split}
\label{eq:grad_L_chain_rule}
\end{equation}

First, we compute the partial derivative of the loss function with respect to an arbitrary intermediate inverse temperature $\beta_j$ ($j = 2, \dots, M-1$). Considering that $\beta_j$ only affects the acceptance rates $A_{j-1,j}$ and $A_{j,j+1}$ of the adjacent pairs, the derivative is given by:
\begin{equation}
\begin{split}
    \frac{\partial f_{\mathrm{uni}}}{\partial\beta_j} = \frac{2}{M-1} 
    \Biggl\{
        &\left(\mathbb{E}[A_{j-1,j}] - \bar{A}\right) \frac{\partial}{\partial\beta_j}\mathbb{E}[A_{j-1,j}] \\
        +&\left(\mathbb{E}[A_{j,j+1}] - \bar{A}\right) \frac{\partial}{\partial\beta_j}\mathbb{E}[A_{j,j+1}] 
    \Biggr\}.
\end{split}
\end{equation}

This expression contains the gradients of the expected acceptance rates, which generally cannot be computed directly. To estimate these terms, we employ the score function method~\cite{Williams1992}. This technique, also known as the log-derivative trick, allows the gradient of an expectation to be expressed as an expectation involving the gradient of the log-probability, without requiring differentiation of the function inside the expectation:
\begin{equation}
    \nabla_{\bm{\theta}} \mathbb{E}[f(X(\bm{\theta}))] = \mathbb{E}[f(X) \nabla_{\bm{\theta}} \log p(X|\bm{\theta})].
\end{equation}
We note that stochastic automatic differentiation (StochasticAD)~\cite{Arya2022,Arya2023} serves as an alternative method for estimating such gradients, which is briefly discussed in Appendix~\ref{app:stocAD}.

Applying the score function method, the term $\partial \mathbb{E}[A_{j-1,j}]/\partial \beta_j$ is calculated as:
\begin{equation}
    \frac{\partial\mathbb{E}[A_{j-1,j}]}{\partial\beta_j} = \mathbb{E}\biggl[\frac{\partial A_{j-1,j}}{\partial\beta_j} + A_{j-1,j}\frac{\partial}{\partial\beta_j}\log p(\bm\sigma;\beta_j)\biggr].
\end{equation}
Here, $p(\bm\sigma;\beta_j)$ is the probability distribution at inverse temperature $\beta_j$. Considering the Boltzmann distribution, $p(\bm\sigma;\beta) \propto \exp(-\beta H(\bm\sigma))$, the derivative of the log-probability becomes:
\begin{equation}
    \frac{\partial}{\partial \beta_j}\log p(\bm\sigma;\beta_j) = \langle H \rangle_{\beta_j} - H(\bm\sigma),
\end{equation}
where $H(\bm\sigma)$ is the energy of state $\bm\sigma$, and $\langle H \rangle_{\beta_j}$ is the expectation of the energy at inverse temperature $\beta_j$.
Consequently, the gradient $\partial \mathbb{E}[A_{j-1,j}]/\partial \beta_j$ can be written analytically as~\cite{MacCallum2018, Hagiwara2024}:
\begin{equation}
    \frac{\partial \mathbb{E}[A_{j-1,j}]}{\partial\beta_j} = \langle H \rangle_{\beta_j}\mathbb{E}[A_{j-1,j}] - \mathbb{E}[H A_{j-1,j}].
\end{equation}
The term $\partial \mathbb{E}[A_{j,j+1}]/\partial \beta_j$ is obtained through a similar calculation.

Next, we evaluate the partial derivative of $\beta_j$ with respect to $\Delta\beta_n$. Since $\beta_j$ is expressed as $\beta_j = \beta_1 + \sum_{n=1}^{j-1}\Delta\beta_n$, we have:
\begin{equation}
    \frac{\partial\beta_j}{\partial\Delta\beta_n} = 
    \begin{cases} 
        1 & (n < j), \\
        0 & (n \ge j).
    \end{cases}
\end{equation}
% Furthermore, the derivative of $\Delta\beta_k$ with respect to $L_i$ is:
% \begin{equation}
%     \frac{\partial\Delta\beta_k}{\partial L_i} = \Delta\beta_i \delta_{ik},
% \end{equation}
% where $\delta_{ik}$ is the Kronecker delta. 
Substituting these results into Eq.~\eqref{eq:grad_L_chain_rule}, the gradient of the loss function with respect to $L_k$ is derived as:
\begin{equation}
    \frac{\partial f_{\mathrm{uni}}}{\partial L_k} = \Delta\beta_k \sum_{j=k+1}^{M-1} \frac{\partial f_{\mathrm{uni}}}{\partial \beta_j} \quad (k=1,\dots,M-2).
\end{equation}
This formula consists entirely of observables estimated by the RXMC simulation.
Due to the constraints fixing the endpoints $\beta_1$ and $\beta_M$, determining $M-2$ elements of $\bm{L}$ automatically fixes the remaining interval. Thus, $\Delta\beta_{M-1}$ is determined by:
\begin{equation}
    \Delta\beta_{M-1} = \beta_M - \beta_1 - \sum_{k=1}^{M-2}\Delta\beta_k.
\end{equation}

This formulation allows us to estimate the gradient directly from sampling data and iteratively update the parameter set $\bm{L}$ using the following update rule:
\begin{equation}
    \bm{L} \leftarrow \bm{L} - \eta \nabla_{\bm{L}}f_{\mathrm{uni}}(\bm\beta),
\end{equation}
where $\eta$ is the learning rate. Algorithm~\ref{alg2} summarizes the proposed iterative procedure based on this update rule.
\begin{algorithm}[H]
    \caption{\small Gradient Descent via Score Function Method}
    \label{alg2}
    \begin{algorithmic}[1]
    \REQUIRE learning rate $\eta$, epochs $T$, batch size $B$
    \item[\textbf{Initialize:}] inverse temperature vector $\bm\beta^{(0)}\!=\!(\beta_1^{(0)}, \dots, \beta_M^{(0)})$, MCMC states $\{\bm\sigma_0^{(i)}\}_{i=1}^M$

    \FOR{$t = 0, 1, \dots, T-1$}
        \STATE Initialize gradient estimate $\bm g^{(t)} = \bm 0$
        \FOR{$b = 1, \dots, B$}
            \STATE Run an independent RXMC simulation with parameters $\bm\beta^{(t)}$.
            \STATE Estimate stochastic gradient $\nabla_{\bm L} f_{\mathrm{uni}}(\bm\beta)|_{\bm\beta=\bm\beta^{(t)}}$ from this $b$-th simulation.
            \STATE $\bm g^{(t)} \leftarrow \bm g^{(t)} + \nabla_{\bm L} f_{\mathrm{uni}}(\bm\beta)|_{\bm\beta=\bm\beta^{(t)}}$
        \ENDFOR
        \STATE $\bm g^{(t)} \leftarrow \bm g^{(t)} / B$
        
        \STATE Update the parameters via stochastic gradient descent: 
        \STATE $\bm L^{(t+1)} \leftarrow \bm L^{(t)} - \eta \bm g^{(t)}$
        \STATE Update $\bm\beta^{(t+1)}$ from $\bm L^{(t+1)}$ ensuring boundary constraints.
    \ENDFOR

    \ENSURE Optimized inverse temperature vector $\bm\beta^{(T)}$.
    \end{algorithmic}
\end{algorithm}
Algorithm~\ref{alg2} employs a batch learning approach. The gradient $\nabla_{\bm{L}}f_{\mathrm{uni}}(\bm{\beta})$ is estimated as an average over a mini-batch of $B$ independent RXMC simulations at each epoch $t$. A gradient estimated from a single simulation where $B=1$ can have high variance due to the stochastic nature of the MCMC process. By averaging over a batch, we obtain a more stable gradient estimate, which helps to stabilize the learning process and improves convergence.

\section{\label{sec:level4}Numerical Simulations}
In this section, we describe the numerical experiments conducted to validate our proposed optimization method. We first detail the general experimental setup common to all models. Then, we present the results for three benchmark models: the two-dimensional (2D) ferromagnetic Ising model, the 2D ferromagnetic XY model, and the 3D Edwards-Anderson spin-glass model.
\subsection{Setup}
We detail the common settings used throughout the following simulations.
The RXMC method involves two alternating processes: local updates and replica exchanges.
For the local updates, we employed the standard Metropolis-Hastings algorithm~\cite{Metropolis1953, Hastings1970} with single-spin flip dynamics.
Following each MCS of local updates, exchange attempts are performed between adjacent replicas.
Specifically, we adopted the standard alternating schedule~\cite{Hukushima1996}, where exchanges between odd-indexed pairs and even-indexed pairs are attempted alternately at every MCS.

As the initial values for the optimization of the inverse temperature vector, we used the geometric progression defined in Eq.~\eqref{eq:geometric}. 
The specific values for the lowest inverse temperature $\beta_1$ and the highest inverse temperature $\beta_M$ are set individually for each target model to sufficiently cover the temperature range of interest, particularly including any critical regions.

In addition to the inverse temperatures, the total number of replicas $M$ is also a crucial parameter. 
While optimizing $M$ to balance the overlap between the Boltzmann distributions of adjacent replicas and computational cost is a critical topic~\cite{Katzgraber2006, Nadler2007b}, our proposed method focuses on optimizing the placement of inverse temperatures for a given fixed $M$. 
Therefore, the total number of replicas $M$ was fixed at $M=20$ for all simulations in this study. The determination of the optimal $M$ is left for future work.

We adopted the variance of the acceptance rates between the adjacent replicas, defined in Sec.~\ref{subsec:3.A}, as the loss function for the optimization. 
To estimate the gradient of this loss function, each independent simulation within a batch consists of an equilibration phase of $N_{\mathrm{eq}}$ MCS followed by a sampling phase of $N_{\mathrm{sam}}$ MCS.
The duration $N_{\mathrm{eq}}$ is chosen to allow the system to reach thermal equilibrium.
The sampling duration $N_{\mathrm{sam}}$ is set to a length sufficient to suppress the statistical variance of the estimated gradient, thereby ensuring stable learning updates.
Specific values for $N_{\mathrm{eq}}$ and $N_{\mathrm{sam}}$ are provided in the respective subsections for each model.

Parameter updates were performed using the Adam optimizer~\cite{Kingma2015} with stochastic gradients estimated from the loss function. 
The hyperparameters were set as follows: learning rate $\eta = 0.005$, epochs $T=300$, and batch size $B=10$. 

As a metric to evaluate the sampling efficiency of the optimized inverse temperatures, we measured the round-trip time. 
For comparison, we also measured the round-trip times for two conventional strategies: a geometric progression as defined in Eq.~\eqref{eq:geometric} and linearly spaced inverse temperatures defined as:
\begin{equation}
    \beta_j = \beta_1 + (\beta_M - \beta_1) \frac{j-1}{M-1} \quad (j = 1, 2, \dots, M).
    \label{eq:inverse_linear}
\end{equation}
Hereafter, we refer to this strategy as ``inverse linear''.

The round-trip times of the proposed and baseline methods were evaluated by $10^7$ MCS for each of the final optimized vectors and the two comparison strategies. 
The distribution of round-trip times observed during these simulations was recorded, and the results were visualized using box plots.

Furthermore, we compared the proposed method with a round-trip time minimization method~\cite{Katzgraber2006} for both the Ising model and the Edwards--Anderson model.
For this comparison, the XY model was not included because we focused on discrete-spin models studied in that work.
This comparison has two objectives.
First, we aim to evaluate the extent to which methods designed to equalize acceptance rates can improve sampling efficiency compared to methods that optimize round-trip time.
Second, we aim to verify the effectiveness of the proposed method from the perspective of computation time.
The round-trip time minimization method requires sufficiently long simulations to evaluate the movement of replicas in temperature space, and thus requires large MCS to optimize the temperature set.
On the other hand, since the proposed method calculates the gradient based on the acceptance rate and energy estimated from short MCS, it has the potential to improve the temperature set with less computational time than methods that measure round-trip time.
Therefore, this comparison verifies the effectiveness of the proposed method in terms of both sampling efficiency and computational cost.
The simulation settings of the round-trip time minimization method were the same as those used for the proposed method, except that the number of swap attempts for the feedback optimization was set to $N_{\mathrm{sw}} = 10^5$.

\subsection{\label{sec:Ising_model}Ising Model}

To verify the proposed method, we start with the 2D ferromagnetic Ising model, which is the simplest spin system. The Hamiltonian $\mathcal{H}_{\mathrm{Ising}}$ of this model is given by:
\begin{equation}
    \mathcal{H}_{\mathrm{Ising}} = -J \sum_{\langle i,j \rangle} \sigma_i \sigma_j,
\end{equation}
where $\sigma_i \!\in\! \{\pm 1\}$ represents the Ising spins on a square lattice with $N=L^2$ total spins ($i=1, \dots, N$). The summation $\sum_{\langle i,j \rangle}$ denotes the sum over all pairs of nearest-neighbor spins. The constant $J$ determines the strength of the interaction between spins, and in this study, we set $J=1$ to consider ferromagnetic interactions. 
The 2D Ising model is known to exhibit a phase transition from an ordered phase to a disordered phase at a finite temperature. Near the critical point, the specific heat diverges and energy fluctuations become maximal. Therefore, this temperature region tends to become a sampling bottleneck in the RXMC method, making the optimization of the temperatures particularly important.

In the simulation, we used a system with a side length of $L=20$ and imposed periodic boundary conditions to suppress finite-size effects. 
The hyperparameters of the initial inverse temperature vector, the geometric progression, were set to $\beta_1=0.1, \beta_M=1.0$ chosen to fully include the critical region. 
Furthermore, we set the number of equilibration steps to $N_{\mathrm{eq}} = 10^2$ MCS and the sampling steps to $N_{\mathrm{sam}} = 10^4$ MCS.

For comparison, we also present the results obtained using the policy gradient method, which was described in Sec.~\ref{subsec:2.C}. 
The following settings were used: the initial temperature vector was the geometric progression, and the initial policy parameter $\bm\theta$ was obtained by transforming this initial vector into the differences of logarithmic inverse temperatures.
The standard deviation of the policy distribution was set as:
\begin{equation}
    \sigma(t) = \frac{0.05}{(\eta + 1)^t},
\end{equation}
where $\eta$ is the learning rate and $t$ is the epoch. This annealing schedule was applied to accelerate convergence. 
Considering that the policy is maximized via gradient ascent, we adopted the negative of the variance of acceptance rates between adjacent replicas as the reward function. 
The optimizer was the same as that used for the proposed method, but the hyperparameters were set to: learning rate $\eta=0.001$, epochs $T=3000$, and a batch size of 10.

\begin{figure}[t!]
    \centering
    \begin{subfigure}[b]{0.48\textwidth}
        \centering
        \begin{tikzpicture}
            \node[anchor=south west,inner sep=0] (image) at (0,0) {
                \includegraphics[width=\textwidth]{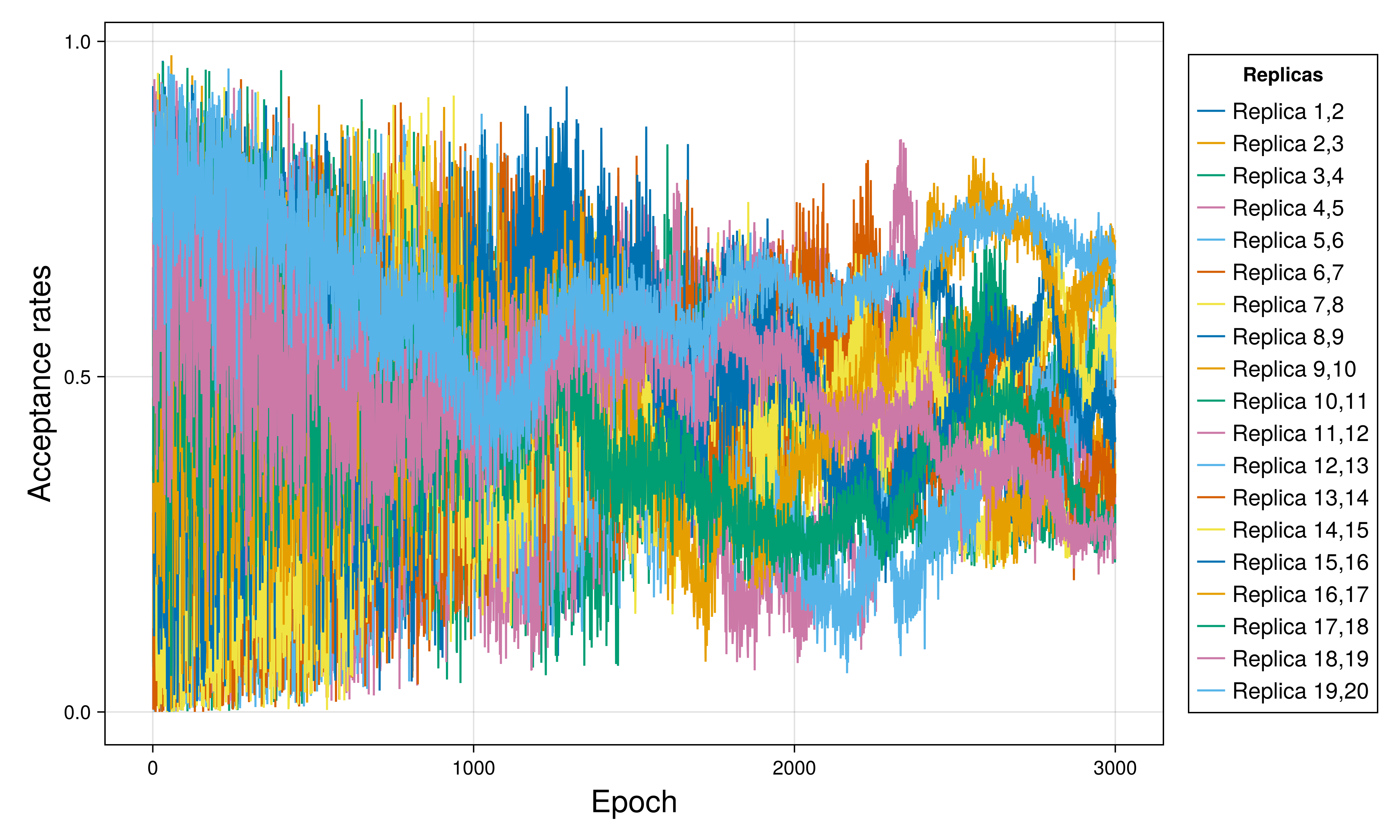}
            };
            \node[anchor=north west, overlay, text=black, xshift=1pt, yshift=9pt] at (image.north west) {(a)};
        \end{tikzpicture}
        % \caption{}
        \label{fig:rates_policy_gradient}
    \end{subfigure}
    \begin{subfigure}[b]{0.48\textwidth}
        \centering
        \begin{tikzpicture}
            \node[anchor=south west,inner sep=0] (image) at (0,0) {
                \includegraphics[width=\textwidth]{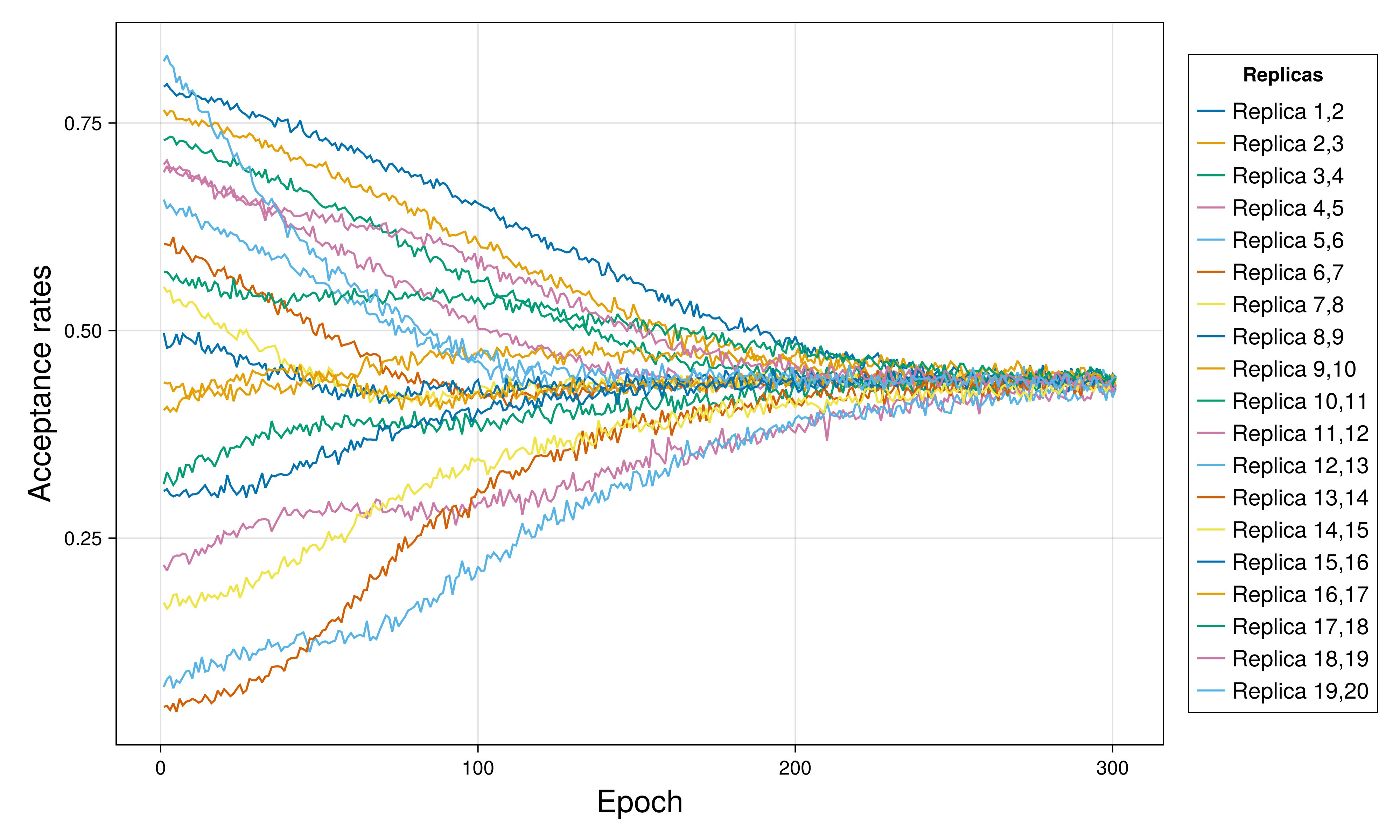}
            };
            \node[anchor=north west, overlay, text=black, xshift=1pt, yshift=9pt] at (image.north west) {(b)};
        \end{tikzpicture}
        % \caption{}
        \label{fig:rates_proposed}
    \end{subfigure}    
    \caption{Evolution of acceptance rates between adjacent replicas during the optimization process for the 2D Ising model ($L=20$). (a) Results using the policy gradient method ($T=3000$). (b) Results using the proposed method ($T=300$).}
    \label{fig:rates}
\end{figure}

\begin{figure}[t!]
    \centering
    \begin{subfigure}[b]{0.48\textwidth}
        \centering
        \begin{tikzpicture}
            \node[anchor=south west,inner sep=0] (image) at (0,0) {
                \includegraphics[width=\columnwidth]{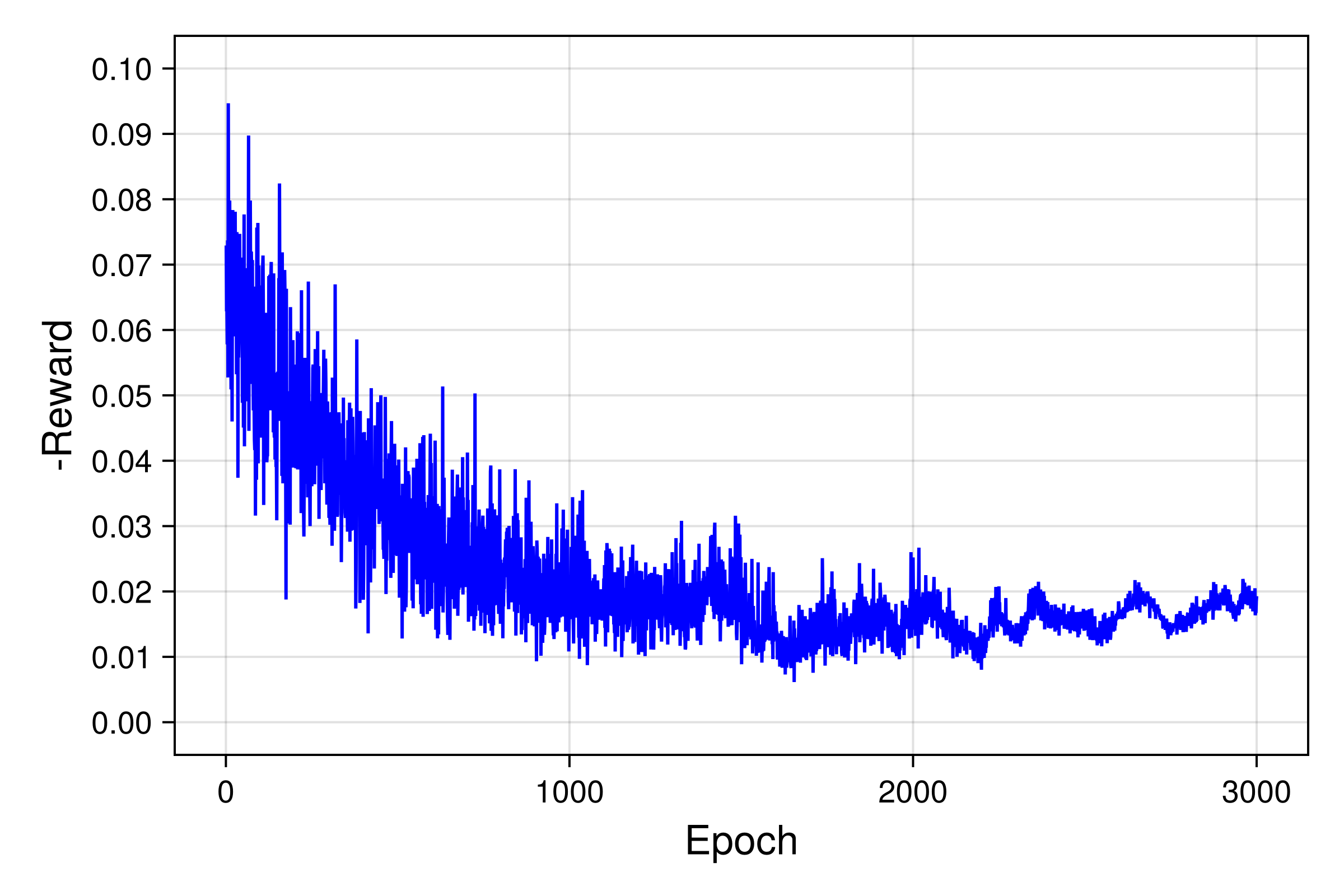}
            };
            \node[anchor=north west, overlay, text=black, xshift=1pt, yshift=9pt] at (image.north west) {(a)};
        \end{tikzpicture}
        % \caption{}
        \label{fig:loss_policy_gradient}
    \end{subfigure}
    \begin{subfigure}[b]{0.48\textwidth}
        \centering
        \begin{tikzpicture}
            \node[anchor=south west,inner sep=0] (image) at (0,0) {
                \includegraphics[width=\columnwidth]{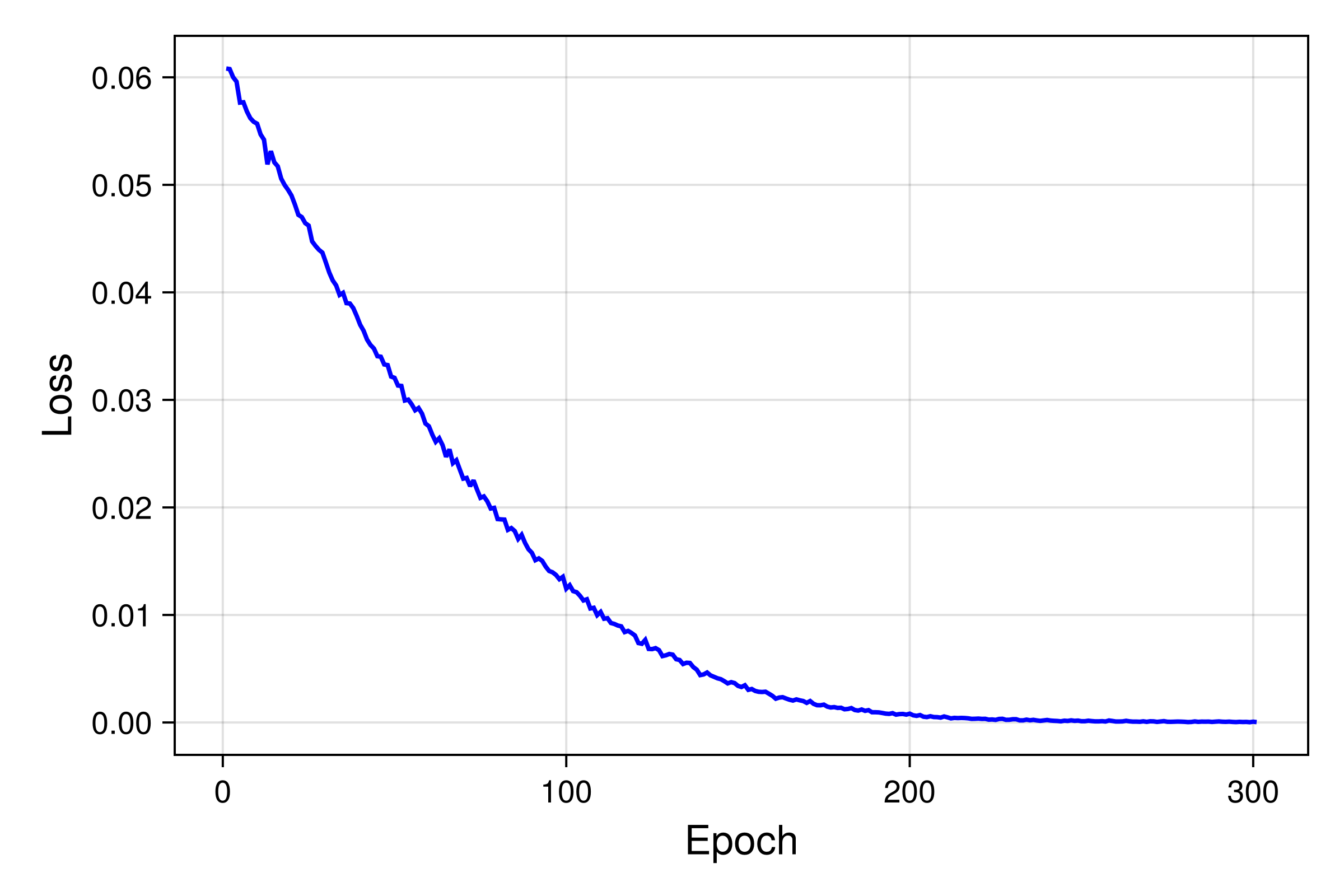}
            };
            \node[anchor=north west, overlay, text=black, xshift=1pt, yshift=9pt] at (image.north west) {(b)};
        \end{tikzpicture}
        % \caption{}
        \label{fig:loss_proposed}
    \end{subfigure}
    \caption{Learning curves for the 2D Ising model ($L=20$). (a) Policy gradient method: The negative reward equivalent to the loss function is plotted against the epoch ($T=3000$). (b) Proposed method: The loss function is plotted against the epoch ($T=300$).}
    \label{fig:losses}
\end{figure}
Figure~\ref{fig:rates} shows the evolution of acceptance rates between adjacent replicas during the optimization steps for both (a) the policy gradient method and (b) the proposed method.
While the acceptance rates for the policy gradient method failed to converge to a single value even after 3000 epochs, the proposed method demonstrated rapid convergence, with the initially dispersed rates smoothly approaching a unified value as the optimization progresses. 
Unlike the optimization using gradient $\nabla_{\bm{\beta}}f_{\mathrm{uni}}$, which is prone to instability due to unconstrained updates (see Appendix~\ref{app:direct_optimization}), our method using the $\log\Delta\bm{\beta}$ reparameterization successfully avoids such instability. 

The stability of our method stems from the global dependency introduced by the variable transformation.
Since each inverse temperature is defined as a cumulative sum of intervals, $\beta_j = \beta_1 + \sum_{n=1}^{j-1} \Delta\beta_n$, a change in a single interval parameter $L_k$ shifts the positions of all subsequent replicas $j > k$.
This is reflected in the chain rule derived in Eq.~\eqref{eq:grad_L_chain_rule}, where the gradient with respect to $L_k$ becomes a summation of gradients from all subsequent replicas.
This means that the update of a local interval is driven not only by its immediate neighbors but by the collective feedback from all replicas at higher inverse temperatures.
Therefore, even if the local acceptance rate drops and the gradient $\partial f / \partial \beta_k$ vanishes locally, the non-vanishing gradients from downstream replicas can propagate back to update $L_k$.
This mechanism prevents the optimization from stagnating in plateaus. Consequently, the optimization proceeds robustly without becoming trapped in poor local minima where acceptance rates drop to zero.

Figure~\ref{fig:losses} shows the learning curves for the policy gradient method and the proposed method. 
In the policy gradient method, the loss fluctuates within the range of $10^{-2}$ to $2\times 10^{-2}$ after 1000 epochs, failing to show clear convergence. 
In contrast, the loss of the proposed method decreases sharply within the first 100 epochs and continues to decrease monotonically thereafter, eventually converging to a value below $10^{-3}$. 
This quantitatively demonstrates that the proposed method successfully achieves its objective of uniform acceptance rates.

We also attempted to optimize the inverse temperature vector using the loss function $f_{\mathrm{log}}(\bm{\beta})$ proposed in \cite{MacCallum2018} within our reparameterization framework.
Since our problem setting fixes both endpoints $\beta_1$ and $\beta_M$, the last interval $\Delta\beta_{M-1}$ is determined dependently as $\Delta\beta_{M-1} = (\beta_M - \beta_1) - \sum_{k=1}^{M-2} \Delta\beta_k$.
In our experiments, we observed that optimization using $f_{\mathrm{log}}$ was highly unstable, similar to the results for protein folding~\cite{MacCallum2018}.
Specifically, the gradients derived from the logarithmic metric aggressively expanded certain intervals, causing the sum of the first $M-2$ intervals to exceed the total range $\beta_M - \beta_1$. 
This resulted in a violation of the boundary constraint $\Delta\beta_{M-1} \ge 0$, leading to a simulation failure.
Therefore, we conclude that the variance-based approach offers superior stability for constrained temperature optimization in discrete spin systems compared to the logarithmic metric.

\begin{figure}[t]
    \includegraphics[width=0.48\textwidth]{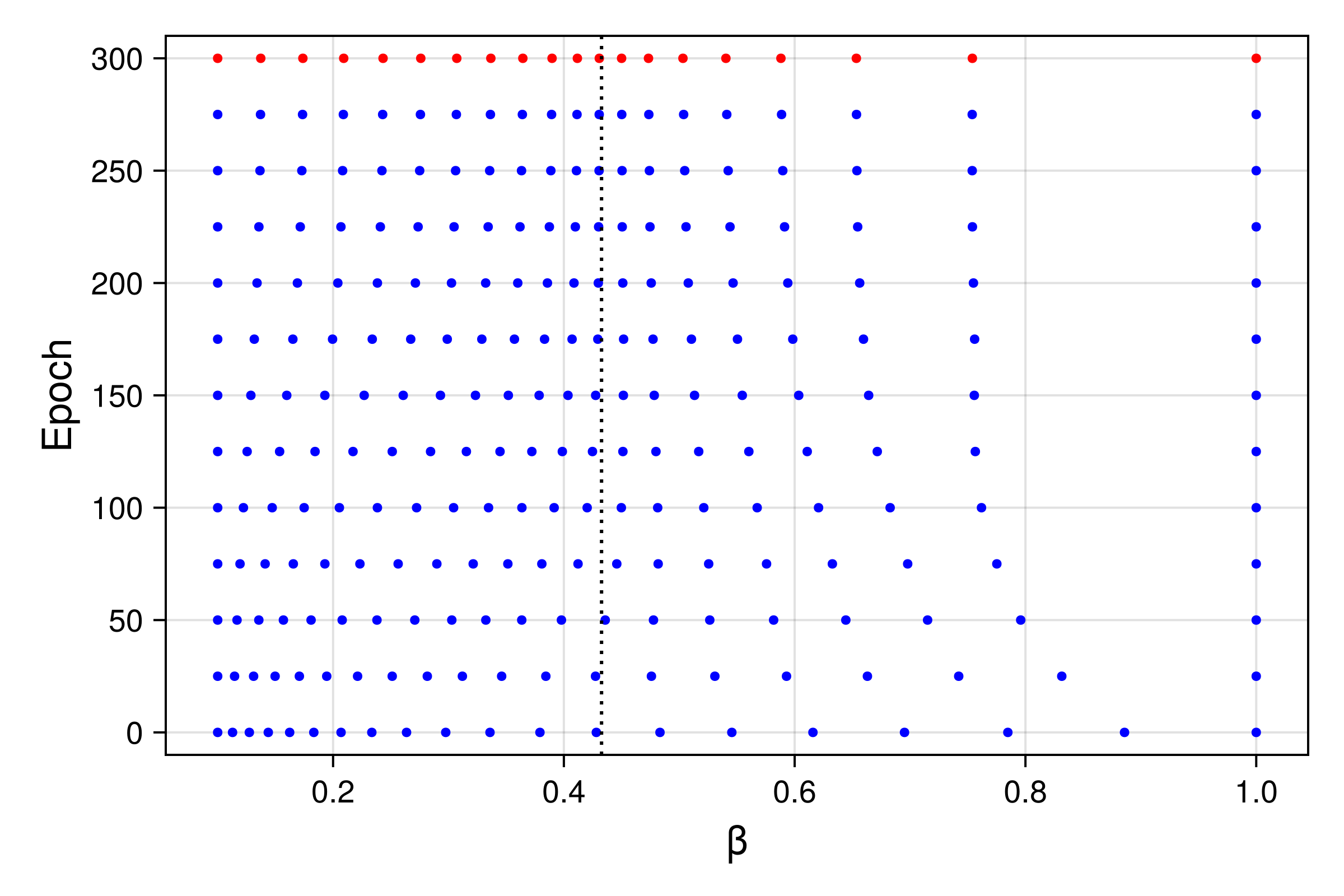}
    \caption{The inverse temperatures at different optimization steps in the 2D Ising model ($L=20$). Starting from the initial geometric progression (epoch 0), the results are shown every 25 steps.
    The dashed line indicates the inverse temperature where the specific heat is maximal.}
    \label{fig:temp_evolution}
\end{figure}

Figure~\ref{fig:temp_evolution} shows the optimization process of the inverse temperatures by the proposed method. 
The horizontal axis represents the inverse temperature $\beta$, and the vertical axis represents the epoch in optimization. 
In the initial state (epoch 0), the inverse temperatures were arranged in a geometric progression. 
As the optimization progresses, each temperature point gradually moves, and it is clearly observed that they eventually cluster in the vicinity of the inverse temperature where the specific heat reaches its maximum.

This behavior is plausible in terms of phase transition. 
Near the critical point, the specific heat exhibits a sharp peak, causing energy fluctuations to become significantly large. 
To maintain a constant acceptance rate between adjacent replicas, it is necessary to ensure sufficient overlap between their Boltzmann distributions. 
This requires making the inverse temperature interval $\Delta\beta$ smaller in the region of high specific heat. 
Our proposed method successfully optimized the temperature intervals to satisfy this physical requirement automatically.
\begin{figure}[t!]
    \centering
    \includegraphics[width=0.48\textwidth]{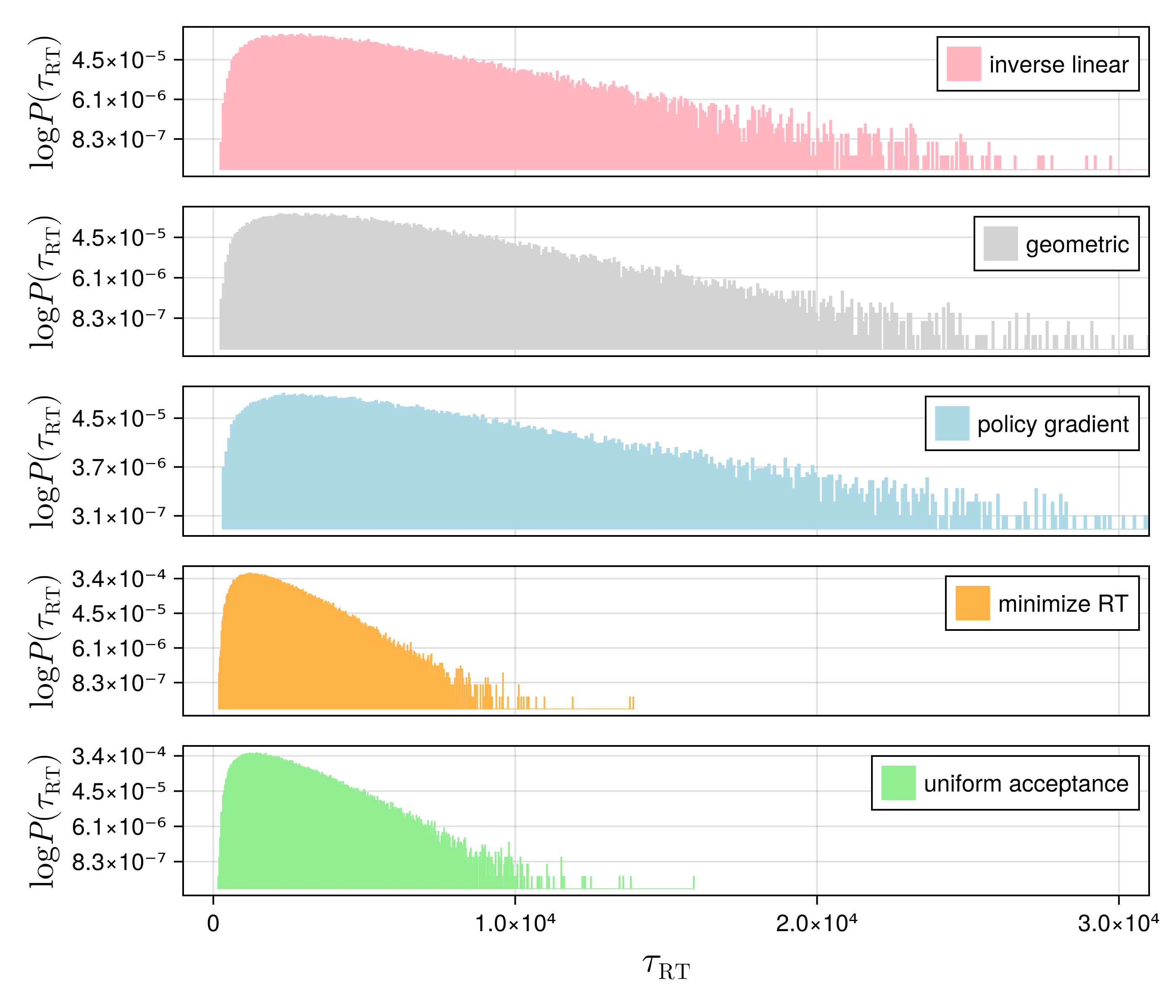}
    \caption{Distribution of round-trip times for the 2D Ising model ($L=20$) under five different temperature selection strategies: from above, inverse linear, geometric progression, the strategy resulting from the policy gradient method, the round-trip time minimization strategy, the uniform acceptance rates strategy optimized using the proposed method. 
    The y-axis is presented on a logarithmic scale. 
    Each density is estimated from an evaluation run of $10^7$ MCS.}
    \label{fig:rtt_dist}
\end{figure}

We attribute the primary reason for the failure of the policy gradient method to the fact that the policy assumes independence among the action components.
In systems exhibiting phase transitions, such as the 2D Ising model, achieving uniform acceptance rates between adjacent replicas necessitates placing inverse temperatures closely spaced near the critical point and widely spaced in other regions, as suggested by Fig.~\ref{fig:temp_evolution}. 
This implies that the action $\bm a$ should not be independent but rather exhibit strong correlations across indices to capture such variation in optimal intervals.
However, since the policy in \cite{Zhao2024} is modeled by a multivariate Gaussian distribution with a diagonal covariance matrix $\sigma^2 \bm{I}$, it assumes that each interval varies independently.
Consequently, during the optimization process, actions significantly distant from the initial policy parameters $\bm\theta_0$ have an extremely low probability of being sampled. 
As a result, the probability of sampling the optimal parameters becomes significantly low, making it difficult to discover the optimal temperature intervals that require exploration across the wide range of scales induced by the phase transition.

\begin{figure}[t!]
    \centering
    \includegraphics[width=0.48\textwidth]{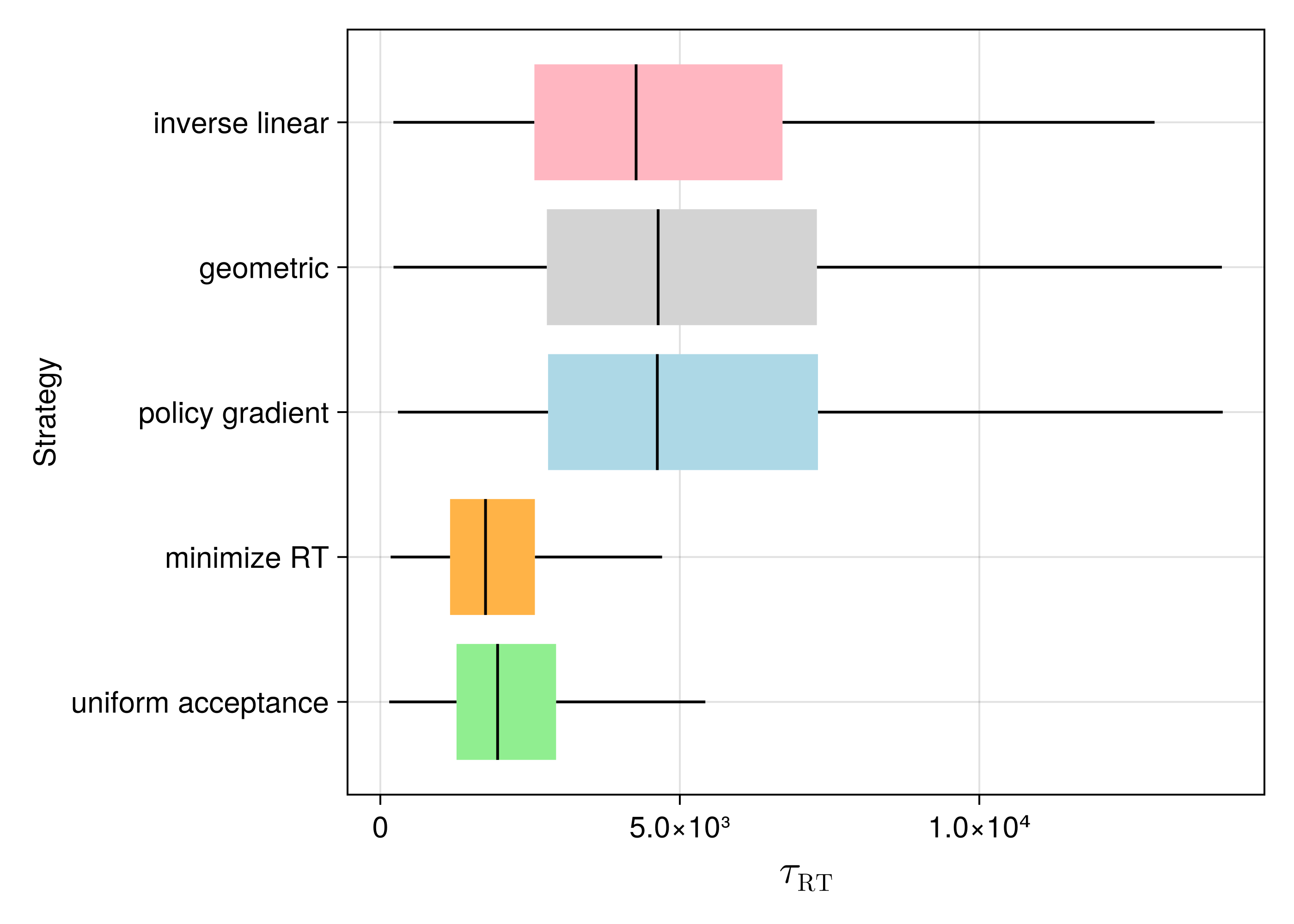}
    \caption{Box plot of the round-trip time distributions for the 2D Ising model ($L=20$) under five strategies. Outliers defined as data points falling beyond 1.5 times the IQR from the box are omitted for clarity.}
    \label{fig:rtt_boxplot_Ising}
\end{figure}

Finally, we evaluated the sampling efficiency using the round-trip time.
Figure~\ref{fig:rtt_dist} shows the distribution of round-trip times observed during $10^7$ MCS runs for each strategy: the proposed method labeled by ``uniform acceptance'', the policy gradient method, the geometric progression, the inverse linear spacing, and the round-trip time minimization method labeled by ``minimize RT''.
This indicates that while the distribution peaks at relatively short durations, it possesses a significant tail where occasional long excursions occur with non-negligible probability. 
Given this long-tailed nature, the mean is highly sensitive to rare, extreme events and may not accurately reflect the typical efficiency. 
Therefore, we adopted the median instead of the mean for evaluation.

Figure~\ref{fig:rtt_boxplot_Ising} shows the box plots of the distribution of round-trip times observed at $10^7$ MCS. 
The centerline of the box represents the median, while the lower and upper edges of the box represent the first and third quartiles, respectively. 
The whiskers represent the range of the data excluding outliers.
The proposed method shows clear improvement over the geometric progression, inverse linear spacing, and policy gradient method.
It achieves a lower median round-trip time, indicated by the center line in the box, and a more compact interquartile range (IQR) than these conventional strategies except for the round-trip minimization method.
Specifically, the proposed method reduced the median round-trip time by 57.7\% compared to the geometric progression.
The round-trip time minimization method achieved the lowest median round-trip time as expected because it directly optimizes the round-trip behavior of replicas.
On the other hand, the proposed method achieved a reduction in round-trip time that rivals that of the round-trip time minimization method, although it does not use round-trip time as its objective function.
This result indicates that the uniformization of the acceptance rate serves as an effective alternative metric for reducing round-trip time. 
In addition, the computational time required for optimization was comparable between the proposed method and the round-trip time minimization method in this experiment.

\subsection{XY Model}
The Hamiltonian $\mathcal{H}_{\mathrm{XY}}$ for the 2D XY model is given by:
\begin{align}
    \mathcal{H}_{\mathrm{XY}} = -J \sum_{\langle i,j \rangle}\cos(\theta_i - \theta_j),
\end{align}
where $\theta_i \in [0, 2\pi)$ represents the angle of the spin vector on site $i$ of a square lattice with $N=L^2$ spins ($i=1, \dots, N$). The term $\sum_{\langle i,j \rangle}$ denotes the sum over all nearest-neighbor spin pairs. We set the interaction strength $J=1$.

The 2D XY model is known to exhibit a Berezinskii-Kosterlitz-Thouless (BKT) transition~\cite{Berezinskii1970, Kosterlitz1973, Kosterlitz1974} at a finite temperature, related to the unbinding of vortex-antivortex pairs. Near the transition temperature, the system exhibits slow dynamics and long correlation lengths, which pose a significant challenge for conventional MCMC sampling. Therefore, optimizing the temperatures for the RXMC method is crucial for efficiently exploring the configuration space around this transition.
\begin{figure}[t]
    \centering
    \includegraphics[width=0.48\textwidth]{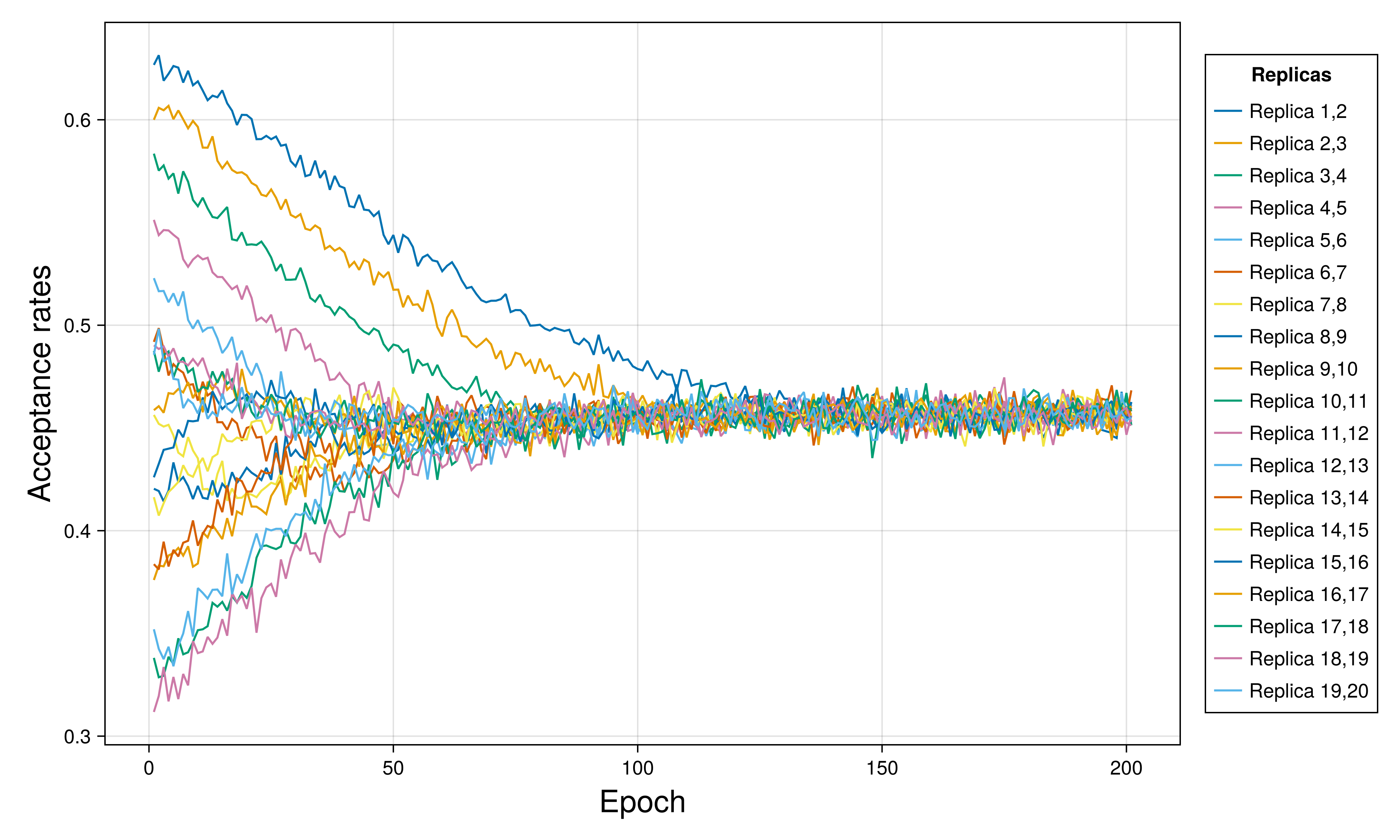}
    \caption{Evolution of acceptance rates between adjacent replicas during the optimization process for the 2D XY model ($L=20$). Although the optimization was performed for 300 epochs, the results are shown only up to epoch 200 because convergence was already achieved by that point.}
    \label{fig:rates_xy}
\end{figure}

In the simulations, we used a system size of $L=20$ and imposed periodic boundary conditions to suppress finite-size effects. 
The hyperparameters of the initial inverse temperature vector, the geometric progression, were set to $\beta_1=0.5, \beta_M=1.5$ chosen to fully include the critical region. 
Furthermore, we set the number of equilibration steps to $N_{\mathrm{eq}} = 10^2$ MCS and the sampling steps to $N_{\mathrm{sam}} = 2\times 10^4$ MCS.

Unlike the experiment of the Ising model, we do not perform comparative experiments with the policy gradient method for this and the subsequent model.
The results presented in Sec.~\ref{sec:Ising_model} indicate that the policy gradient method failed to converge. 
This failure suggests a fundamental limitation of this method for systems exhibiting phase transitions. 
Consequently, it is expected that similarly inadequate performance will be observed when applying it to the 2D XY model or the 3D Edwards--Anderson model.
% Therefore, we omit these comparisons and focus exclusively on validating the effectiveness and versatility of our proposed method in the remainder of this paper.

Figure~\ref{fig:rates_xy} shows the evolution of acceptance rates between adjacent replicas. 
Similar to the Ising model, the initially dispersed acceptance rates rapidly converge to a unified value, demonstrating the effectiveness of the method.
Consistent with the Ising model results, the optimization proceeded smoothly without stagnation.

\begin{figure}[t]
    \centering
    \includegraphics[width=0.48\textwidth]{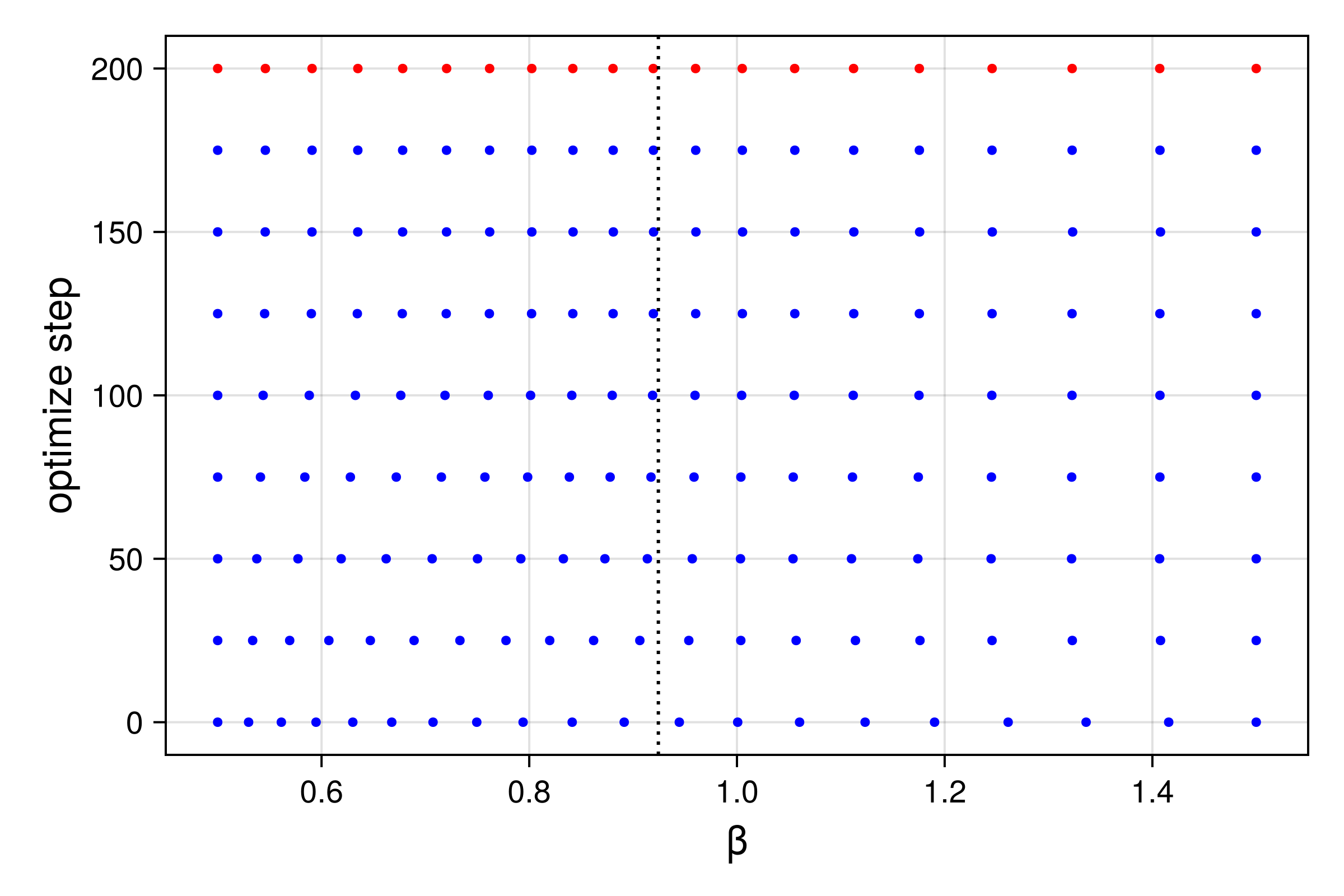}
    \caption{The inverse temperatures at different optimization steps in the 2D XY model ($L=20$). Starting from the initial geometric progression (epoch 0), the results are shown every 25 steps. The dashed line indicates the inverse temperature where the specific heat peaks.}
    \label{fig:temp_evolution_xy}
\end{figure}

Figure~\ref{fig:temp_evolution_xy} shows the optimization process of the inverse temperatures. 
Similar to the behavior observed in the Ising model, the inverse temperatures which started from a geometric progression (epoch 0) autonomously cluster as the optimization progresses. 
They concentrate in the vicinity of the inverse temperature where the specific heat exhibits its broad peak. 
This result again confirms the ability of the proposed method to automatically identify the critical region where sampling is most difficult, corroborating the findings from the Ising model.

\begin{figure}[t]
    \centering
    \includegraphics[width=0.48\textwidth]{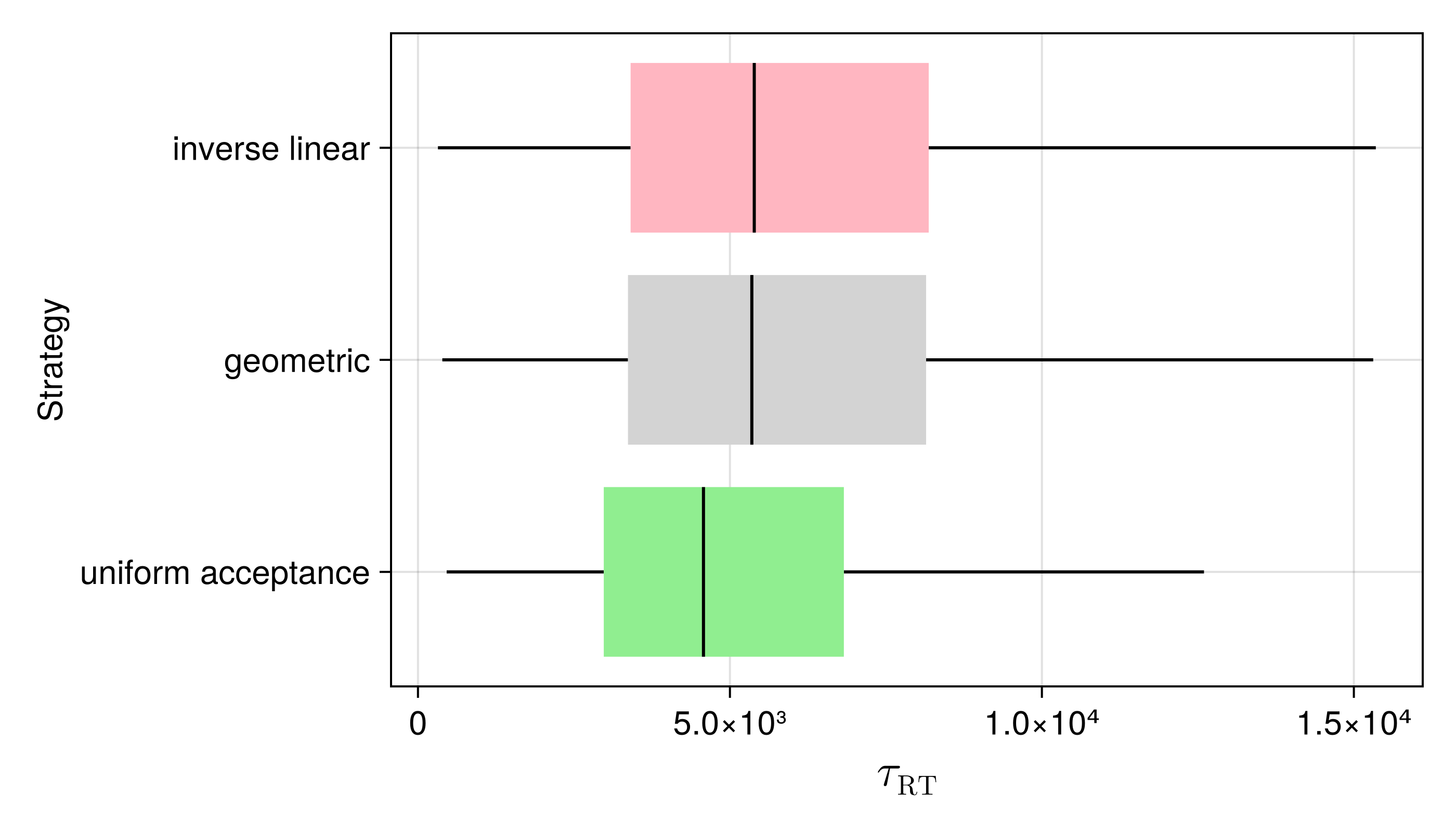}
    \caption{Box plot of the round-trip time distributions for the 2D XY model ($L=20$) under three strategies: from above, inverse linear, geometric progression, the uniform acceptance rates strategy optimized using the proposed method, with outliers removed for clarity.}
    \label{fig:rtt_boxplot_xy}
\end{figure}

Figure~\ref{fig:rtt_boxplot_xy} presents a box plot comparing the round-trip time distributions for the uniform acceptance rates achieved by the proposed method against the geometric progression and inverse linear strategies. 
Consistent with the Ising model results, the proposed method achieved the lowest median round-trip time and the most compact IQR. 
Specifically, the proposed method reduced the median round-trip time by 14.5\% compared to the geometric progression.
This result is particularly significant as it demonstrates that our proposed method is effective not only for discrete spin systems like the Ising model but also for continuous variable systems like the XY model.
This successful application to a system with continuous degrees of freedom highlights the versatility and robustness of the proposed framework, confirming its applicability regardless of whether the state variables are discrete or continuous.

\subsection{Edwards-Anderson Model}
Following the verification on ferromagnetic systems, we extend our analysis to the 3D Edwards-Anderson (EA) model~\cite{Edwards1975}. 
This model serves as a canonical example of spin glasses, allowing us to test the efficacy of the proposed method in a more challenging scenario involving quenched disorder and a complex energy landscape.
The Hamiltonian $\mathcal{H}_{\mathrm{EA}}$ of the 3D EA model is given by:
\begin{align}
    \mathcal{H}_{\mathrm{EA}} = - \sum_{\langle i,j \rangle} J_{ij} \sigma_i \sigma_j,
\end{align}
where $\sigma_i \!\in\! \{\pm 1\}$ represents an Ising spin on a cubic lattice with $N=L^3$ spins ($i=1, \dots, N$). The sum $\sum_{\langle i,j \rangle}$ runs over all nearest-neighbor pairs. The interactions $J_{ij}$ are quenched random variables drawn independently from a Gaussian distribution with zero mean and unit variance.

The sampling efficiency of the RXMC method becomes critically dependent on the temperatures for this model, especially near the spin-glass transition point where the specific heat becomes large. 
The resulting large energy fluctuations cause a significant reduction in the overlap between the Boltzmann distributions of adjacent replicas, creating a bottleneck. 
Thus, careful optimization of the temperatures is essential for overcoming this bottleneck and ensuring an efficient random walk in the temperature space.

In the simulations, we used a system size of $L=8$ and imposed periodic boundary conditions. 
The hyperparameters of the initial inverse temperature vector, the geometric progression, were set to $\beta_1=0.1, \beta_M=1.5$ chosen to fully include the critical region. 
Furthermore, we set the number of equilibration steps to $N_{\mathrm{eq}} = 10^2$ MCS and the sampling steps to $N_{\mathrm{sam}} = 2\times 10^4$ MCS.

\begin{figure}[t]
    \centering
    \begin{tikzpicture}
        \node[anchor=south west,inner sep=0] (image) at (0,0) {
            \includegraphics[width=0.48\textwidth]{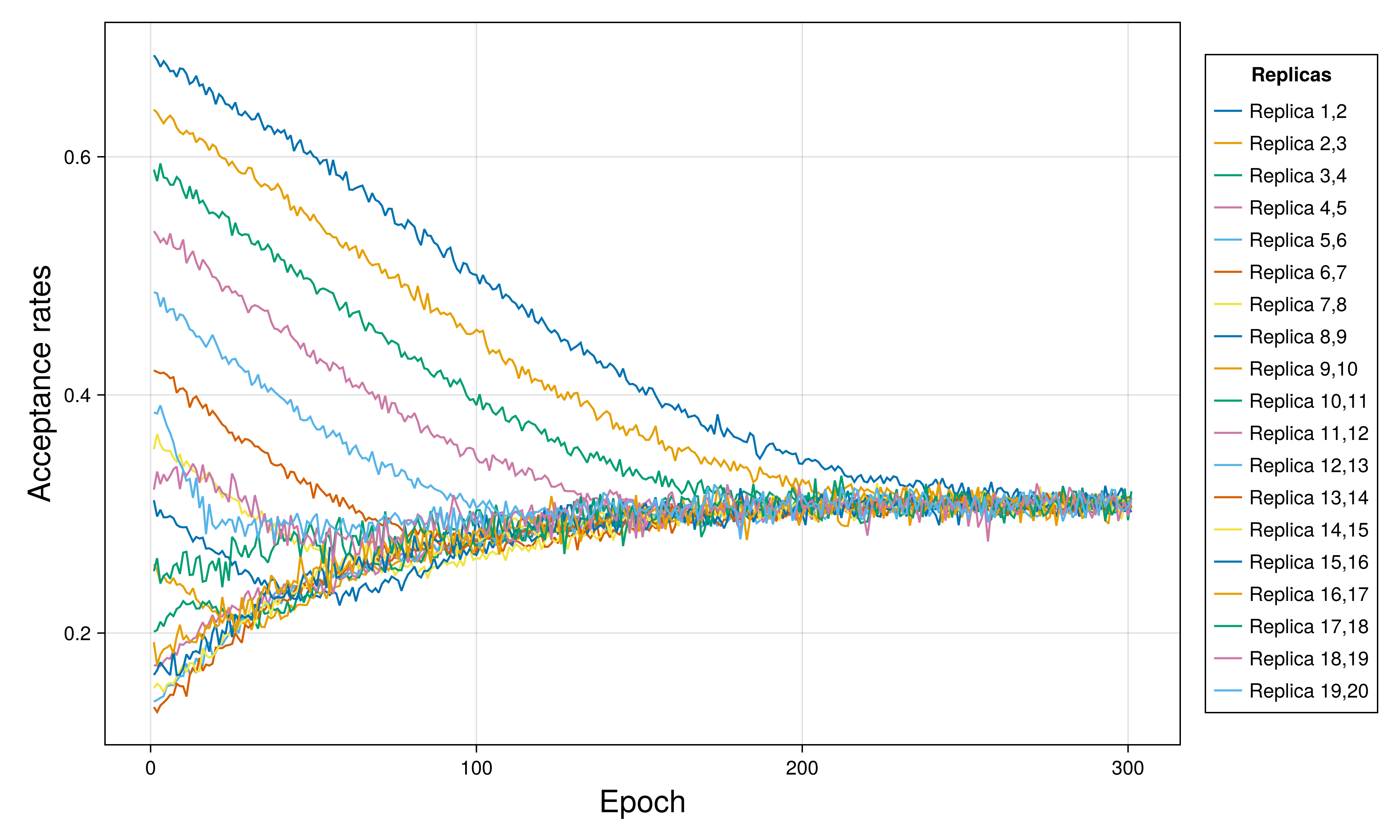}
        };
    \end{tikzpicture}
    \caption{Evolution of acceptance rates between adjacent replicas during the optimization process for the 3D EA model ($L=8$).}
    \label{fig:rates_ea}
\end{figure}

\begin{figure}[t]
    \centering
    \includegraphics[width=0.48\textwidth]{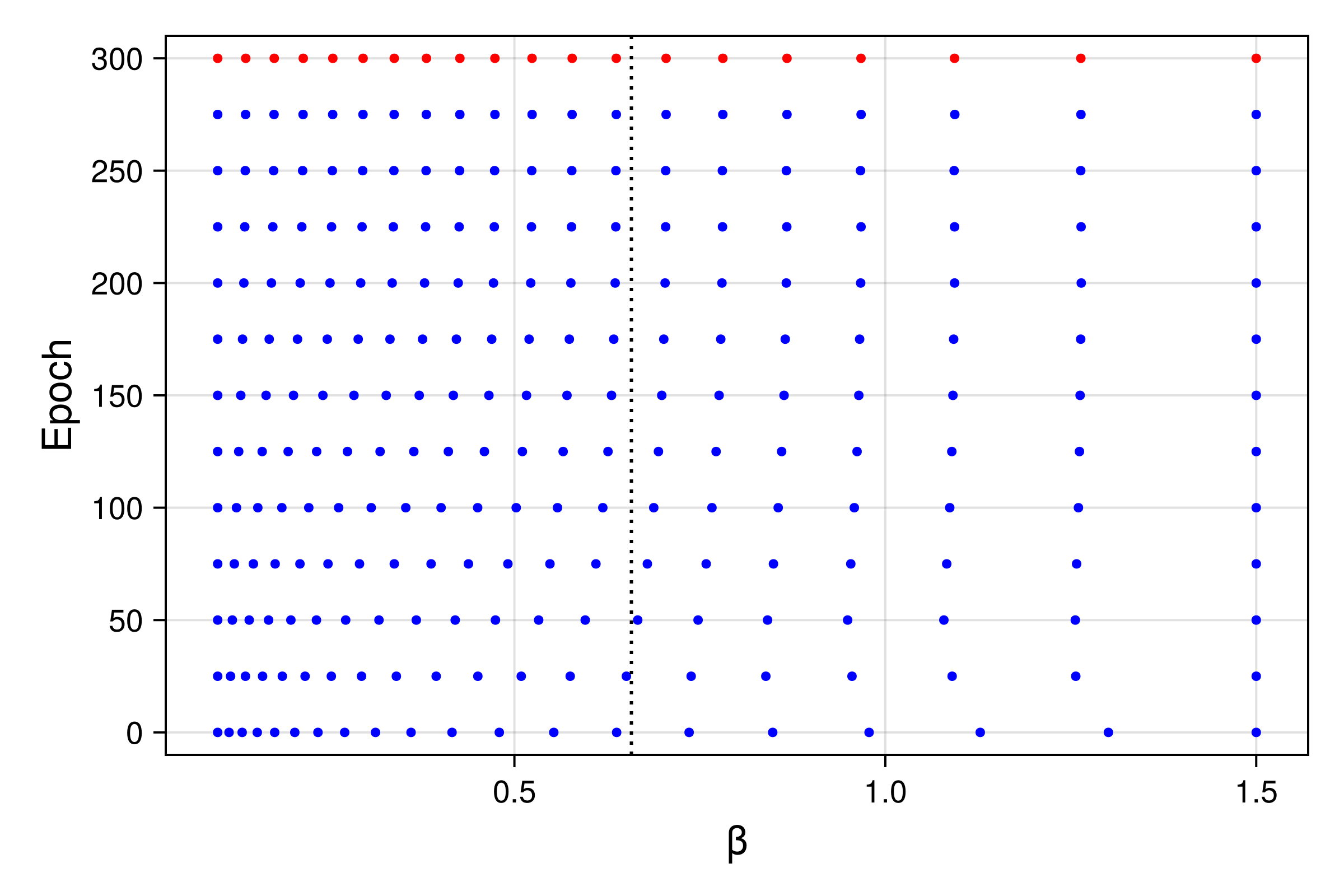}
    \caption{The inverse temperatures at different optimization steps in the 3D EA model ($L=8$). Starting from the initial geometric progression (epoch 0), the results are shown every 25 steps. The dashed line indicates the inverse temperature where the specific heat is maximal.}
    \label{fig:temp_evolution_ea}
\end{figure}

\begin{figure}[t]
    \centering
    \includegraphics[width=0.48\textwidth]{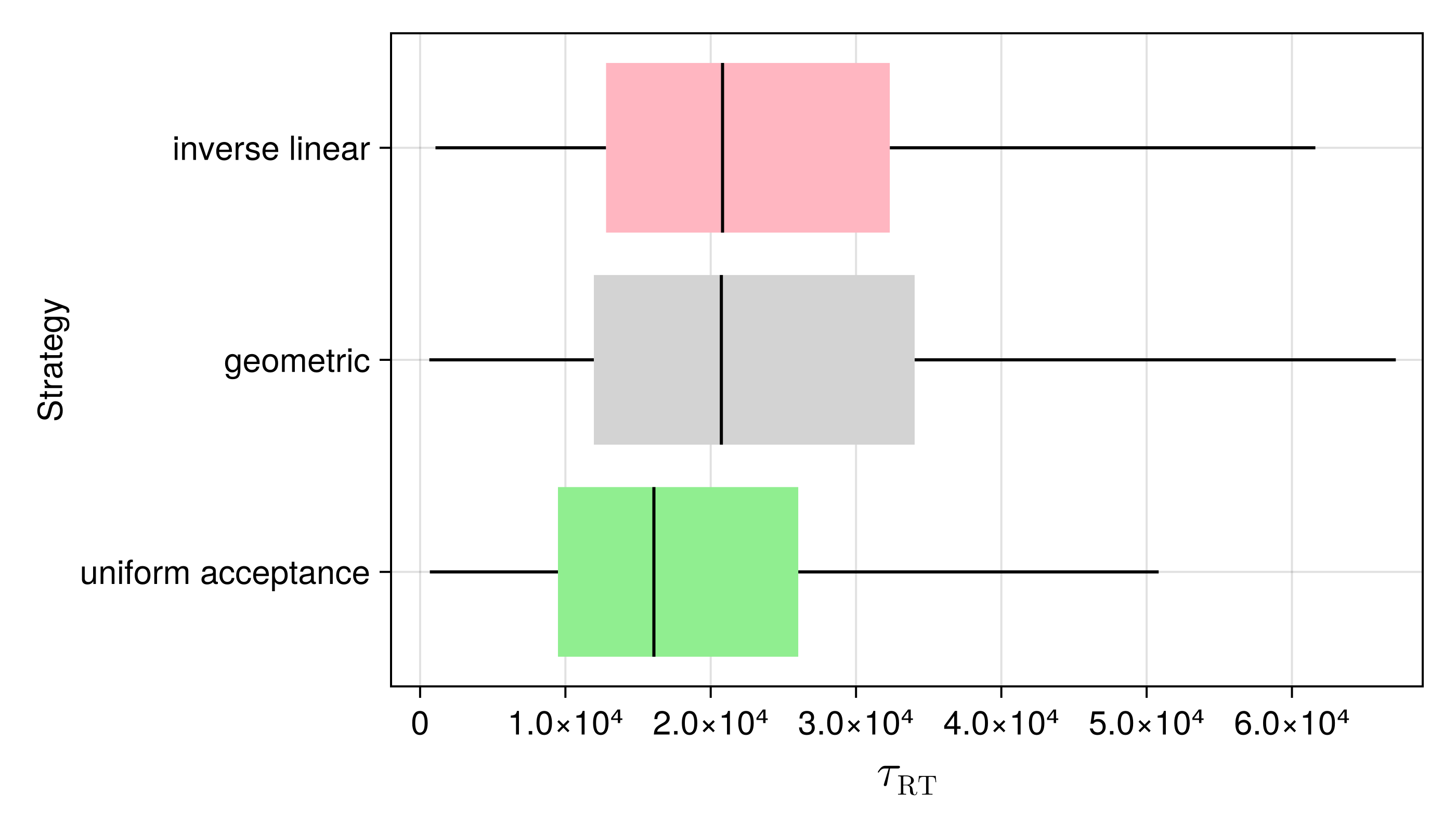}
    \caption{Box plot of the round-trip time distributions for the 3D EA model ($L=8$) under three strategies, with outliers removed for clarity. Result for the round-trip time minimization strategy is not shown because it did not converge within the prescribed number of MCS.}
    \label{fig:rtt_boxplot_ea}
\end{figure}

Despite the presence of quenched disorder, the proposed method demonstrates robust performance comparable to that observed in ferromagnetic models. 
As shown in Fig.~\ref{fig:rates_ea}, the initially dispersed acceptance rates converge smoothly to a uniform value. 
Correspondingly, as shown in Fig.~\ref{fig:temp_evolution_ea}, the inverse temperatures autonomously adapt to the temperature dependence of the specific heat, concentrating them near the critical region where the specific heat peaks. 
This optimization leads to a significant enhancement in sampling efficiency as illustrated in Fig.~\ref{fig:rtt_boxplot_ea}, the proposed method achieved the lowest median round-trip time and the most compact distribution among the tested strategies. 
Specifically, it reduced the median round-trip time by 22.4\% compared to the geometric progression.

We also applied the round-trip time minimization method to the same EA model.
For the particular set of quenched interactions used in this study, however, the round-trip time minimization method did not reach convergence within MCS comparable to that required for the proposed method.
Since the round-trip time minimization method relies on the global motion of replicas over the entire temperature range, its evaluation becomes costly when relaxation is slow.
In contrast, the proposed method updates the temperature set from acceptance rates between adjacent replicas and does not require direct observation of completed round trips during optimization.
This difference explains the practical advantage of the proposed method for slowly relaxing disordered systems.

These results confirm the efficacy of our gradient-based framework even for spin-glass systems characterized by complex energy landscapes.

Regarding the application to the EA model, it should be noted that the optimization in this experiment was performed for a single instance of quenched interactions. 
For spin glasses, the energy landscape depends on the interaction instance, and an inverse temperature vector optimized for one instance is not necessarily optimal for other instances.
To verify it, we independently generated 100 interaction instances that were different from the instance used in the optimization from the Gaussian distribution described earlier, and measured the acceptance rates between adjacent replicas using the optimized inverse temperatures to investigate the instance dependence of the obtained inverse temperatures.
\begin{figure}[t]
    \centering
    \includegraphics[width=0.48\textwidth]{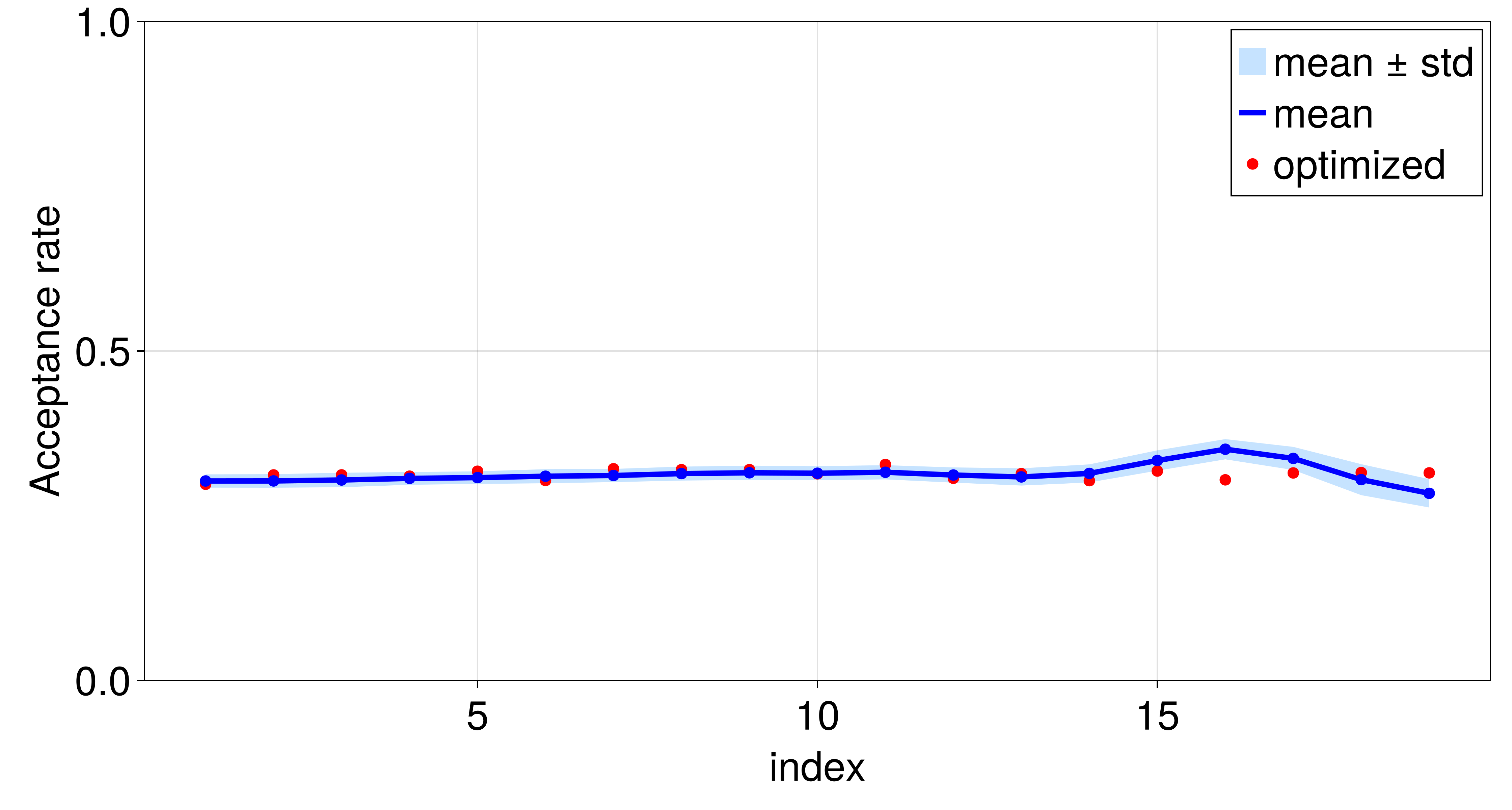}
    \caption{The acceptance rates between adjacent replicas when an inverse temperatures optimized for a single interaction instance is applied to other interaction instances. 
    The index $i$ corresponds to the subscript $i$ of the acceptance rate $A_{i,i+1}$. 
    The line represents the instance-averaged acceptance rate, and the bands represent the standard deviation. 
    The red dots represent the acceptance rates when applied to the interaction instance used in the optimization.}
    \label{fig:compare_instance}
\end{figure}

Figure~\ref{fig:compare_instance} shows the instance-averaged acceptance rates between adjacent replicas and their standard deviations.
Although complete uniformity was not achieved, the average acceptance rate remained roughly the same between each pair of adjacent replicas, and the standard deviation was small.
This result suggests that the inverse temperature vector optimized for a single instance can still provide reasonable acceptance rate uniformity for other interaction instances.
On the other hand, this result does not imply that the inverse temperatures obtained from a single instance are optimal for any instance.
A more practical strategy would be to construct a single inverse temperature vector shared among multiple instances~\cite{Katzgraber2006}.
This could be achieved by performing the optimization over multiple different instances and minimizing an instance-averaged loss function.
A detailed investigation of such an instance-averaged optimization approach is left for future work.

In summary, our simulations demonstrated the effectiveness of the proposed method across three benchmark spin systems: the 2D Ising model, the 2D XY model, and the 3D EA model.
For every model, the optimization of the inverse temperatures based on uniform acceptance rates led to a reduction in round-trip times compared to conventional strategies.
The comparison with the round-trip time minimization method further indicates that, while round-trip time minimization is effective for relatively simple systems such as the Ising model, it can become computationally demanding for slowly relaxing spin-glass systems.
These results support the usefulness of the proposed gradient-based framework for improving the sampling efficiency of RXMC simulations.

\section{\label{sec:level5}Conclusion}
In this study, we proposed a refined online temperature optimization algorithm by extending an existing gradient-based framework~\cite{MacCallum2018}.
A key feature of our approach is the introduction of a reparameterization technique that optimizes the logarithm of the inverse temperature differences rather than the inverse temperatures themselves.
This formulation explicitly imposes the monotonic ordering of inverse temperatures and fixed boundary constraints, effectively resolving the instability and constraint violations inherent in existing parameter optimization methods.
By using the variance of acceptance rates as the loss function, our method achieved uniform acceptance rates across the benchmark spin systems and showed more stable optimization behavior than the policy gradient method.
The optimized temperatures reduced the round-trip times of replicas, indicating improved sampling efficiency.
Furthermore, the optimized temperatures were concentrated near the critical region where the specific heat peaks in all benchmark systems.
This indicates that the proposed method can identify bottleneck regions from acceptance rates between adjacent replicas, without manual tuning or prior knowledge of the model.

The proposed framework is not limited to replica-exchange Monte-Carlo simulations.
Since the loss functions used in this study are defined through acceptance rates between adjacent replicas, the same idea can in principle be applied to replica-exchange molecular dynamics~(REMD).
In REMD, the local sampling dynamics is replaced by molecular dynamics, whereas the exchange step and its acceptance rate have the same form as in RXMC.
Therefore, the proposed gradient-based temperature optimization can be incorporated into REMD without changing the basic formulation.

Several other important considerations remain.
First, applications to continuous particle systems such as systems with Lennard--Jones potential are an important direction for future work.
Such systems are widely used as benchmark models for molecular simulations.
Therefore, applying the proposed method to these systems is necessary to examine whether the proposed temperature optimization framework remains effective beyond spin systems.

Second, systems with strong first-order phase transitions require further investigation.
Although the proposed method is in principle applicable to these systems, its numerical validation for strong first-order transitions requires sufficiently long simulations.
Near a first-order transition point, phase coexistence and large free-energy barriers can make the acceptance rates between adjacent replicas extremely small.
The resulting long relaxation time makes it difficult to obtain reliable statistics for gradient estimation from short MCS runs.
Therefore, the verification of the proposed method for systems with strong first-order transitions is left for future work.

Third, regarding the loss function, our comparative analysis showed that not all differentiable metrics are suitable for online optimization.
Indeed, the metric $f_{\mathrm{log}}$ in \cite{MacCallum2018} exhibited unstable behavior, which complicates the determination of convergence criteria and limits practical reliability.
This issue can also arise for any loss function that has high estimation variance or sensitivity to boundary conditions.
Therefore, while our gradient-based framework offers flexibility in the choice of loss function, the selection of a metric that ensures both statistical stability and reasonability is essential for successful optimization.

Finally, addressing the scalability of the gradient estimator is an important direction for future research.
The score function method adopted in this study is known to suffer from large variance, which typically increases with the system size.
This high variance may destabilize the learning process or require small learning rates, thereby slowing down convergence in large-scale simulations.
To mitigate this issue, improvements to the score function estimator, such as the implementation of control variates or advanced baselines, are promising approaches~\cite{Mohamed2020}.
Alternatively, lower-variance estimation methods, such as the differentiation frameworks discussed in Appendix~\ref{app:stocAD}, could improve the stability of the proposed method and broaden its applicability to larger and more complex systems.

\section*{Acknowledgements}
The authors would like to thank K.~Hukushima and Y.~Nishikawa for their fruitful discussions and helpful comments. This work is partly supported by JSPS Grant-in-Aid for Early-Career Scientists Grant Numbers 19K14613, 26K14992 (ST) and 25K21297 (SA).

\appendix

\section{\label{app:direct_optimization}Necessity of Reparametrization}

To demonstrate the necessity of the proposed reparameterization strategy, we conducted a comparative experiment applying gradient descent directly to the inverse temperature vector $\bm{\beta}$ without enforcing constraints.
In this experiment, the learning rate was set to a conservative value of $\eta=0.001$ to prevent violations of the monotonic ordering, while all other experimental settings followed those described in Sec.~\ref{sec:Ising_model}.
\begin{figure}[t!]
    \centering
    \begin{subfigure}[b]{0.48\textwidth}
        \centering
        \begin{tikzpicture}
            \node[anchor=south west,inner sep=0] (image) at (0,0) {
                \includegraphics[width=\textwidth]{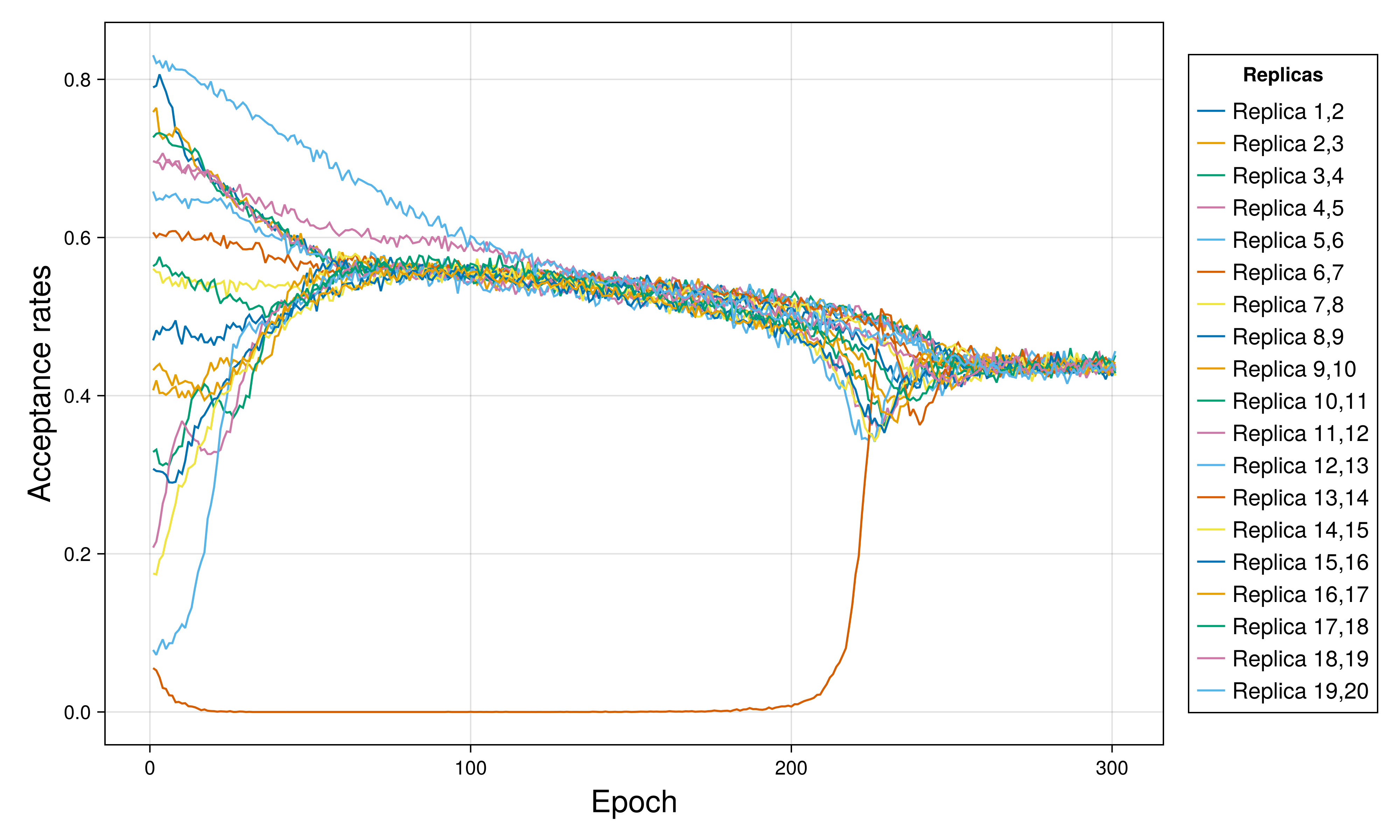}
            };
            \node[anchor=north west, overlay, text=black, xshift=1pt, yshift=9pt] at (image.north west) {(a)};
        \end{tikzpicture}
        % \caption{}
        \label{fig:direct_acc}
    \end{subfigure}
    \begin{subfigure}[b]{0.48\textwidth}
        \centering
        \begin{tikzpicture}
            \node[anchor=south west,inner sep=0] (image) at (0,0) {
                \includegraphics[width=\textwidth]{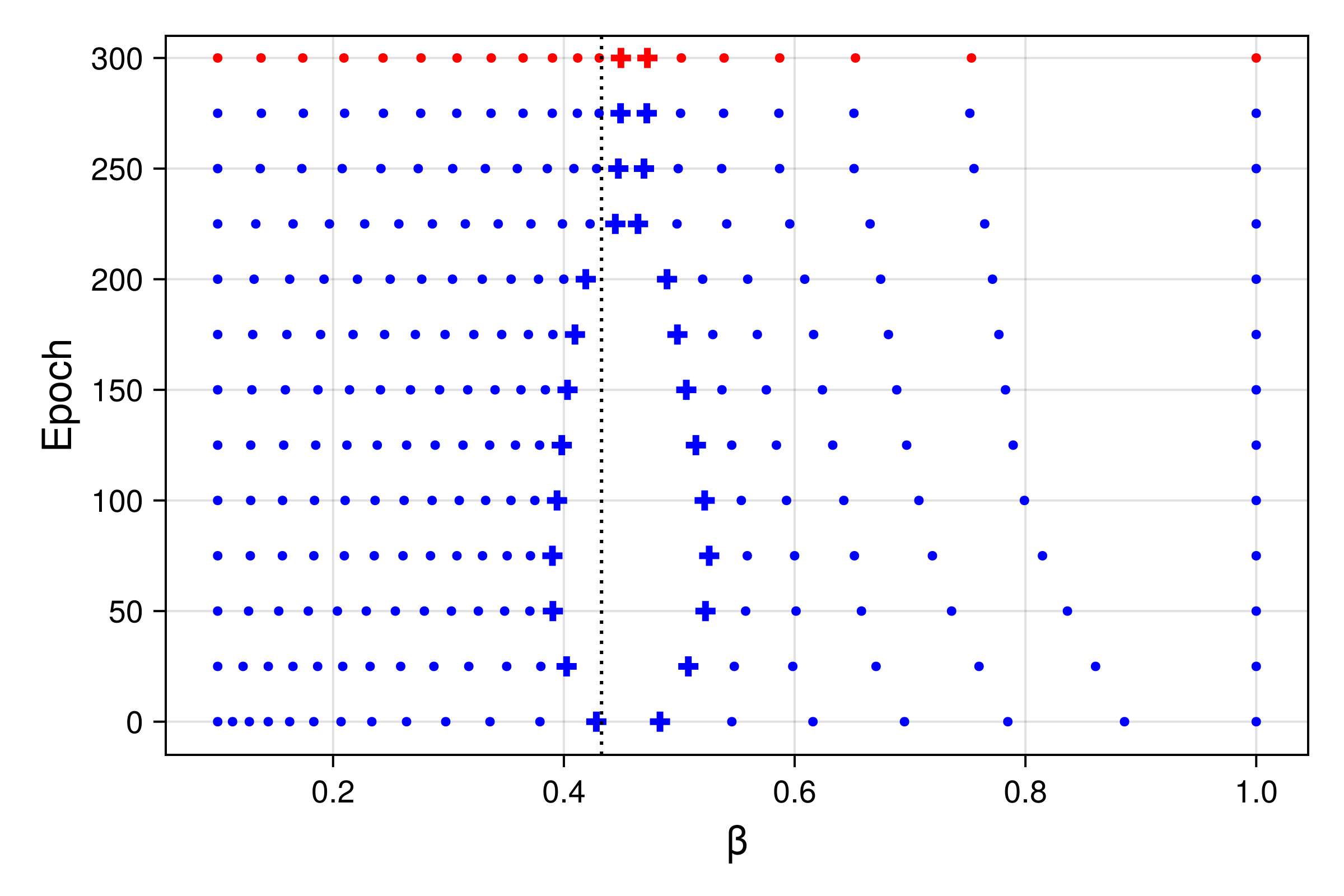}
            };
            \node[anchor=north west, overlay, text=black, xshift=1pt, yshift=9pt] at (image.north west) {(b)};
        \end{tikzpicture}
        % \caption{}
        \label{fig:direct_ladder}
    \end{subfigure}
    \caption{Unconstraint optimization results using gradient descent on $\nabla_{\bm{\beta}}f$ in the 2D Ising model ($L=20$). 
    (a) Evolution of acceptance rates between adjacent replicas over 300 epochs. The rate for the pair between replicas 13 and 14 drops to zero, causing stagnation for over 200 epochs. 
    (b) Evolution of inverse temperatures. Starting from a geometric progression (epoch 0), snapshots are plotted every 25 steps. 
    The dashed line indicates the inverse temperature where the specific heat is maximal. 
    The $+$ symbol represents the 13th and 14th replicas.}
    \label{fig:direct_opt_results}
\end{figure}

Figure~\ref{fig:direct_opt_results}(a) shows the evolution of acceptance rates between adjacent replicas.
The acceptance rate for the specific pair of replicas 13 and 14 drops to near zero immediately and remains stagnant for over 200 epochs.
This stagnation corresponds to the behavior observed in Fig.~\ref{fig:direct_opt_results}(b), which visualizes the evolution of the inverse temperatures.
A gap appears between the corresponding inverse temperatures near the region of maximal specific heat.

This phenomenon illustrates the instability inherent in the gradient descent method using $\nabla_{\bm{\beta}} f_{\mathrm{uni}}$.
Depending on the initial temperatures, updates may cause the inverse temperatures near the critical region to move away from its region, resulting in an acceptance rate between corresponding replicas that approaches zero.
Consequently, the accuracy of gradient estimation decreases, making it difficult to obtain updates that correct the excessively widened inverse temperature interval. 
Therefore, regions where the acceptance rate approaches zero not only reduce sampling efficiency but can also lead to failure in temperature optimization.
\begin{figure}[t!]
    \centering
    \begin{subfigure}[b]{0.45\textwidth}
        \centering
        \begin{tikzpicture}
            \node[anchor=south west,inner sep=0] (image) at (0,0) {
                \includegraphics[width=0.92\columnwidth]{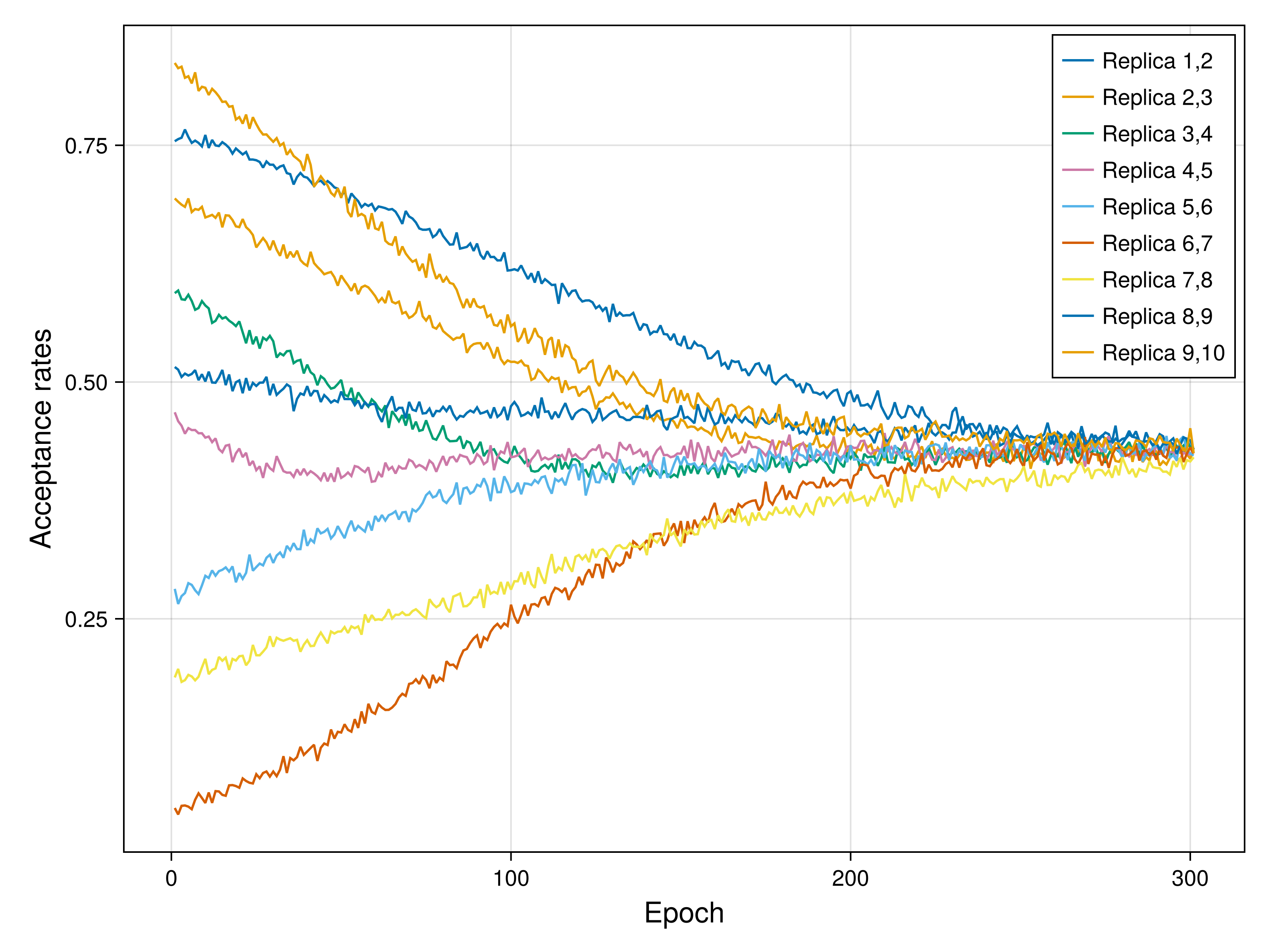}
            };
            \node[anchor=north west, overlay, text=black, xshift=1pt, yshift=9pt] at (image.north west) {(a)};
        \end{tikzpicture}
        % \caption{}
        \label{fig:sad_rates}
    \end{subfigure}
    \begin{subfigure}[b]{0.48\textwidth}
        \centering
        \begin{tikzpicture}
            \node[anchor=south west,inner sep=0] (image) at (0,0) {
                \includegraphics[width=\columnwidth]{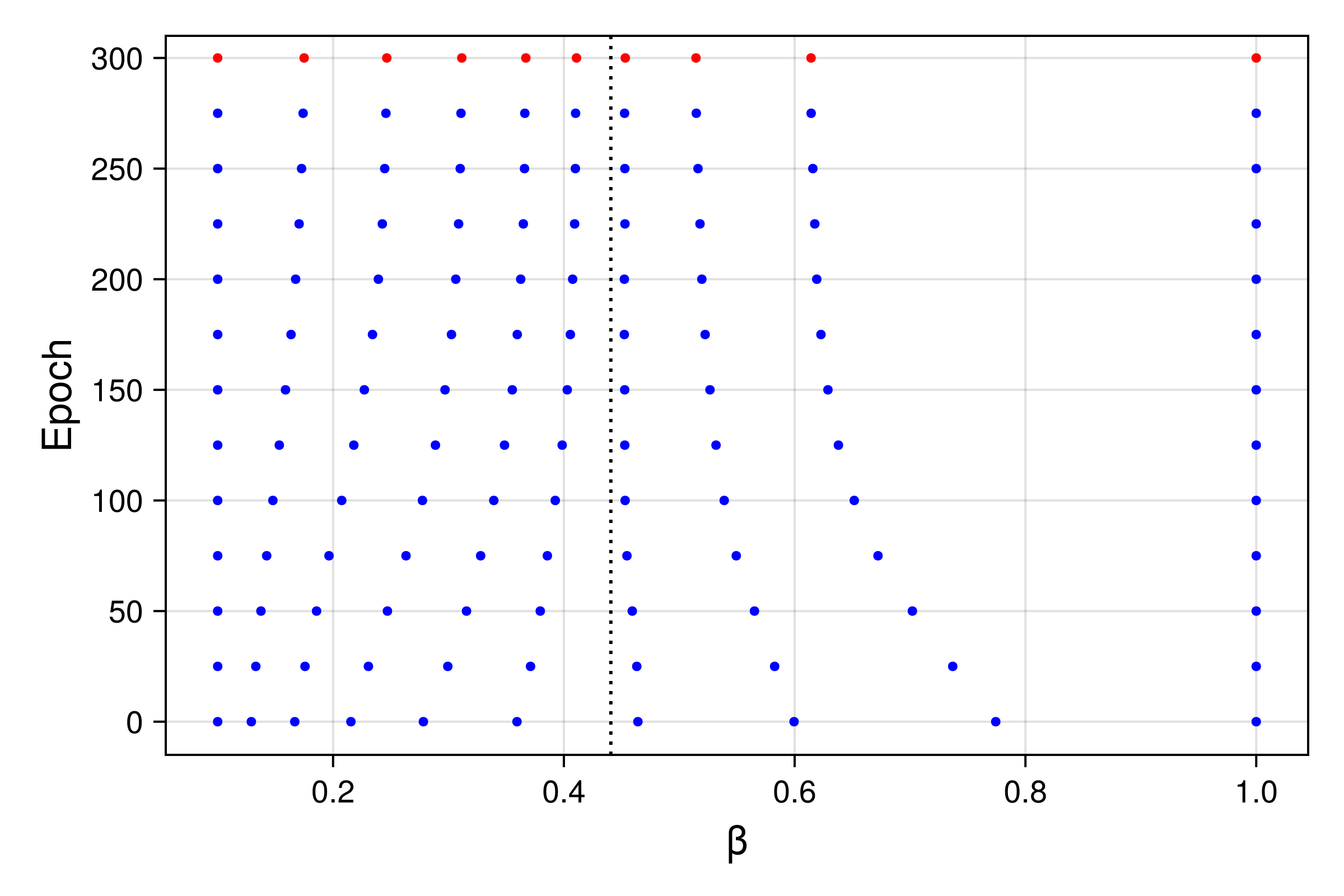}
            };
            \node[anchor=north west, overlay, text=black, xshift=1pt, yshift=9pt] at (image.north west) {(b)};
        \end{tikzpicture}
        % \caption{}
        \label{fig:sad_evolution}
    \end{subfigure}
    \caption{Optimization results using StochasticAD for the 2D Ising model ($L=10$). (a) Evolution of acceptance rates between adjacent replicas over 300 epochs. (b) Evolution of inverse temperatures. Starting from a geometric progression (epoch 0), snapshots are plotted every 25 steps. The dashed line indicates the inverse temperature of maximal specific heat.}
    \label{fig:stocAD_results}
\end{figure}

In this experiment, the optimization process escaped this stagnation around epoch 230. 
This escape is thought to have occurred not because enough gradient information was obtained from the 13th and 14th replicas, but because sampling was functioning in the surrounding temperature regions. 
In fact, as shown in Fig.~\ref{fig:direct_opt_results}(b), the acceptance rates between adjacent replicas away from the critical region are converging toward a uniform value as optimization progresses.
This suggests that gradient in the surrounding critical regions were effectively estimated and updates based on them continued in temperature intervals where there was no stagnation.
On the other hand, in systems such as spin glasses or those exhibiting first-order phase transitions, relaxation may become significantly slower not only near the critical point but also in the surrounding temperature regions. 
In such cases, it becomes difficult to obtain reliable statistics even in the vicinity of the stagnation region, and the effectiveness of resolving stagnation through updates from the surrounding regions is likely to be weakened. 
Therefore, once a region where the acceptance rate approaches zero appears in such systems, optimization is highly likely to lead to failure.

These results underscore the importance of our proposed reparameterization.

% \clearpage
% \onecolumngrid
\section{\label{app:stocAD}StochasticAD}

In this work, we employed the score function method to estimate the gradient of the loss function. As an alternative, we also investigated the use of stochastic automatic differentiation (StochasticAD), another powerful framework for estimating the gradients of expectations, particularly for programs involving discrete random variables~\cite{Arya2022, Arya2023}. Unlike the log-derivative trick used in the score function method, StochasticAD extends the reparameterization trick~\cite{Kingma2014} to discrete variables by incorporating concepts from finite-difference methods~\cite{Asmussen2007}. A key advantage of this approach is that it often yields lower-variance gradient estimators~\cite{Arya2023}.

To conduct a comparative study, we replaced the gradient estimation module of our framework with StochasticAD, implemented using the \texttt{StochasticAD.jl} package in Julia~\cite{JuliaLang}. We then applied this modified framework to the same temperature optimization problem for the 2D Ising model. This experiment was performed on a smaller system with a side length of $L=10$ and $M=10$ replicas, while all other hyperparameters, such as the learning rate and optimizer, remained identical to those in the main experiment.

As shown in Fig.~\ref{fig:stocAD_results}(a), the StochasticAD-based approach also successfully optimizes the inverse temperature vector, achieving uniform acceptance rates. 
Moreover, as illustrated in Fig.~\ref{fig:stocAD_results}(b), the final optimized temperatures were qualitatively consistent with the results obtained by the score function method, such as concentrating inverse temperatures near the critical point. 
However, we observed a significant practical drawback: the computational overhead of StochasticAD was substantial. 
The total processing time for the optimization was considerably longer than that of the score function method. This computational cost presents a major bottleneck for applying this method to larger system sizes.

% \clearpage
% \twocolumngrid
% \FloatBarrier
% \clearpage
\nocite{*}
\bibliography{reference}
\end{document}